\documentclass[aps, prd, 12pt, preprint, a4paper, twoside, showpacs, nobalancelastpage]{revtex4-1}

\usepackage{multirow, amsmath, newfloat, bm}
\usepackage{algorithm, algcompatible}
\usepackage{newfloat, color}

\usepackage[mathscr]{euscript} 
\usepackage[caption=false]{subfig}
\usepackage{graphicx}
\usepackage[dvipsnames]{xcolor}
\usepackage{hyperref}
\hypersetup{
  colorlinks=true,
  citecolor=ForestGreen,
  urlcolor= WildStrawberry,
  allbordercolors={0 0 0},
  pdfborderstyle={/S/U/W 1}
}

\def \Hz	{\mathrm{Hz}}

\def \rlist	{\mathscr{U}}
\def \klist	{\mathscr{K}}
\def \tlist	{\mathscr{T}}
\def \msun	{\mathrm{M}_\odot}
\def \tfrs 	{\mathrm{TaylorF2RedSpin}}
\def \aligo	{\mathrm{aLIGOZeroDetHighPower}} 
\def \flow	{f_{\mathrm{low}}}
\def \fhigh	{f_{\mathrm{high}}}
\def \Mmin	{{\mathcal{M}}_{\mathrm{min}}}

\begin{document}

%\title{Efficient template placement for gravitational wave searches from NS-BH compact binary coalescence}
\title{A hybrid geometric-random template placement algorithm for gravitational wave searches from compact binary coalescences}

\author{Soumen Roy}			\email[]{soumen.roy@iitgn.ac.in}
\author{Anand S. Sengupta}	\email[]{asengupta@iitgn.ac.in}
\author{Nilay Thakor}     \email[]{thakor.nilaysinh@iitgn.ac.in}
\affiliation{Indian Institute of Technology Gandhinagar\\Gujarat 382355, India.}

%%\date{\today}

\begin{abstract}

Astrophysical compact binary systems consisting of neutron stars and blackholes are an important class of gravitational wave (GW) sources for advanced LIGO detectors.  Accurate theoretical waveform models from the inspiral, merger and ringdown phases of such systems, are used to filter detector data under the template based matched filtering paradigm.
%in order to determine their presence in detector noise. 
An efficient grid over the parameter space at a fixed minimal match has a direct impact on the overall time taken by these searches. 
%While the deterministic geometrical template placement algorithm using $A_2^*$ lattice was used extensively in initial-LIGO era for non-spinning searches, it has not been explored for higher dimensions. More recently, the stochastic template placement algorithm with its ease of implementation and scalability to higher dimensions has been used by the LIGO Scientific Collaboration (LSC) but is considered sub-optimal. 
We present a new hybrid geometric-random template placement algorithm for signals described by two masses and one spin magnitude parameters. Such template banks could potentially be used in GW searches from binary neutron stars and neutron star-blackhole systems. The template placement is robust and is able to automatically accommodate curvature and boundary effects with no fine tuning. We also compare these banks against vanilla-stochastic template banks and show that while both are equally efficient in the fitting-factor sense, the bank sizes are $\sim 25 \%$ larger in the stochastic method. Further, we show that the generation of the proposed hybrid banks can be sped-up by nearly an order of magnitude over the stochastic bank. Generic issues related to optimal implementation are discussed in detail. These improvements are expected to directly reduce the computational cost of gravitational wave searches.

\end{abstract}


\maketitle 

\section{Introduction}

%Gravitational waves (GW) are ripples in spacetime caused (to leading order) by the time-dependent quadrapole deformations of mass-energy  distributions. These waves propagate outwards at the speed of light and are a key prediction of Einstein's geometric theory of gravity better known as General Relativity \cite{EinsteinGTR}. While the 1974 discovery of the Hulse-Taylor binary pulsar PSR B1913+16 system \cite{HulseTaylor-1974} provided the first evidence of the {\it emission} of gravitational waves from compact binary systems; their direct detection has been an outstanding challenge of 21st century science. 
The direct detection of gravitational waves (GW) was vigorously pursued by the LIGO Scientific Collaboration (LSC) consisting of several hundred scientists which culminated with the discovery of the first gravitational wave event GW150914 \cite{gw150914} in the twin advanced-LIGO (advLIGO) \cite{advancedLIGO} detectors. This event was determined to have been caused by the  inspiral and merger of a spinning binary blackhole system of component masses 
$(36, \;29) \msun$ located nearly 1.3 billion light years away from Earth. It was also the first observational evidence for the existence of stellar mass blackholes. Subsequently, advLIGO detectors also detected a second event GW151226 \cite{gw151226} from the inspiral and merger of lighter compact objects. These detections mark the transition to the era of gravitational wave astronomy.

Several other kilometer scale detectors are under upgradation or construction at present around the globe - these include the French-Italian advanced Virgo detector \cite{advancedVIRGO-2015} and the Japanese cryogenic KAGRA detector \cite{kagra-2012, kagra-2013}. In India, the proposal for the advanced LIGO-India detector \cite{indigo-proposal-2011, indigo-unnikrishnan} has been approved and is expected to be built over the next few years. The network of these advanced detectors is expected to improve their overall science potential and herald a new wave in astronomy with the potential to observe the very early Universe and compliment information gathered by electromagnetic observations.

Binary blackholes and neutron stars are considered one of the most promising sources for the advanced terrestrial detectors. Precise theoretical model waveforms for GW emitted from the inspiral, merger and ringdown phases of such compact binary coalescences (CBC) are now available. These models are parametrized by the system's {\it intrinsic} properties such as the component masses, spin etc. and {\it extrinsic} properties such as sky location, distance to the source, time at coalescence etc. Accurate theoretical models allow the use of matched filtering technique to search for weak GW signals buried in detector noise. 
As the signal parameters are not known \textit{a priori}, one filters the data using a set of expected signals spanning the deemed parameter space. Each one of these expected signal corresponds to a single point in the parameter space, and are collectively known as the template bank. Coverage of the full range of search parameters using a finite grid of discrete points leads to an inevitable loss in the signal-to-noise ratio (SNR) which can be controlled by fixing the minimal match of the bank. The latter is often decided by striking a balance between desired detection efficiency and computational cost of carrying out the search. 

Advanced detectors are reaching unprecedented sensitivities at low frequencies. Over the last few years, the development of theoretical spinning waveform models have also reached a mature stage.  The combined effect of these factors is that one now needs to search over a significantly larger volume compared to initial LIGO era, in a parameter space spanning three or more dimension - posing new challenges for data analysis, which include devising efficient grid placement strategies. The practical issue of optimal template placement in matched-filter searches is an important open optimization problem \cite{coverart, Prix-2009}.

%The construction of template bank, or in other words, gridding the parameters space is a key problem in gravitational wave data analysis. 
There are at least two approaches to solving the problem of efficient template placement - (a) via deterministically placed points through the tessalations of a geometrical lattice \cite{Cokelaer-2007, Babak-2006} or (b) via stochastically placed points by choosing them at random \cite{Harry-2009} over the parameter space. The geometric method requires the metric over the signal manifold and has been used extensively for CBC searches over 2D mass parameters by arranging the grid points in a hexagonal lattice. A variant of the geometrical placement method for aligned 
spin CBC systems has been explored \cite{Brown-2012} for the TaylorF2 \cite{Buonanno-2009, Poisson-1995} signal model. 
% However its extension to arbitrary signal models in 3D remain an open problem. 
Lack of availability of metric and also the intricate fine tuning required to avoid uncovered regions arising from variations in curvature across the parameter space also make
it difficult to generalize the geometric placement methods to higher dimension.
% However, it has not been successfully implemented in higher dimensions because of lack of availability of metric and also the intricate fine tuning required to avoid uncovered regions arising from variations in curvature across the parameter space.
 
The stochastic template bank is constructed from random proposals drawn from a uniform distribution over the deemed parameter space that are accepted as a new template point only if the new proposal is far (in minimal match sense) from existing templates in the bank. Such banks are easy to implement, robust and can work even without the explicit knowledge of the metric on the signal manifold by using brute force match calculations. This approach has been demonstrated to be more space efficient than a square lattice in 2D but less efficient than a hexagonal lattice. Such banks can be calculated for higher dimensions as well \cite{Ajith-2014}.

An earlier attempt to combine the geometric and stochastic methods \cite{Capano-2016} by seeding a stochastic bank with a pre-fabricated geometric bank has also been tried and has been demonstrated to improve the efficiency marginally.

We present a new hybrid algorithm for template placement in 3D (two mass components and one reduced spin magnitude) parameter space for gravitational wave searches from compact binary coalescence by combining the efficiency of optimal geometrical placement and the robustness and ease of stochastic placement algorithm. This geometric-random bank placement method uses a local truncated octahedral lattice to place the templates and requires the metric over the signal manifold.

\subsection*{Plan of the Paper:}
The paper is organized as follows: 

In Section \ref{sec:sigm} we recapitulate the fundamentals of templated matched filtering technique used for gravitational wave searches from CBC and review the definition of the metric on the signal manifold with a view to set the notation used in the paper and also define key terms. Numerous papers on template placement strategies have been written over the last few years and many new ideas have emerged \cite{Brown-2012, Capano-2016, Harry-2009, Ajith-2014}: we provide a concise review of these efforts for the benefit of the reader. In particular, we elucidate the stochastic algorithm by casting it in two different ways which are algorithmically equivalent. This sets the stage for Section \ref{sec:algorithm}, where we present a new hybrid geometric-random template placement algorithm and give a detailed explanation of the issues concerning its optimal implementation. We argue that by construction this new method can not be less space efficient than the vanilla-stochastic placement method. We also construct an explicit template bank for neutron star-blackhole searches using this new algorithm and compare it against a vanilla-stochastic template bank, noting the improvement in overall bank size as well as the time taken to generate the bank. The $\tfrs$ signal model \cite{Ajith-2011b} was used for this purpose. The template banks presented in this work have been calculated using the $\aligo$ \cite{aLIGOZeroDetHighPower} noise sensitivity curve for advLIGO.

In Section \ref{sec:testOfAlgorithm}, the template banks generated by the hybrid method are tested and validated against the vanilla stochastic bank. We present the fitting factor results using $50,000$ signal injections using $\tfrs$  and TaylorF2 aligned spin waveform models using software implemented in  the \texttt{LALApps} package of the LIGO Algorithm Library \cite{lalsuite} and show that the two banks are nearly identical in efficiency. We also compare the hybrid bank against a $A_3^*$ lattice seeded vanilla-stochastic bank and show that the former is more efficient.

Finally, in Section \ref{sec:Conclusions}, we summarize the main results and make some comments related to several key issues related to the new method, indicating a possible way for extending it to higher dimensions. 

\section{Metric on the signal manifold}
\label{sec:sigm}

In this section we shall quickly summarize the basics of  template based matched filtering technique used in GW searches from CBC. A basic assumption in matched filtering based searches is that the astrophysical GW signal buried in the detector noise is faithfully represented by the signal model used in the search over the range of search parameters.

The signal manifold ${\bm{\lambda}}$ is the set of all possible GW signals $h({\vec{\lambda}}) \equiv h(t; {\vec{\lambda}})$ characterized by the parameter vector $\vec{\lambda}$. It is customary to represent the corresponding frequency domain signal as $\tilde{h}(\vec{\lambda}) \equiv \tilde{h}(f; \vec{\lambda})$. The detector output $s(t)$ consists of  detector noise $n(t)$ and a possible gravitational wave signal of unknown parameters. The additive noise throws the signal out from this manifold to the space of all possible functions.  In order to find the point in  signal manifold closest to the detector output $s(t)$, the latter is projected over the signal manifold by calculating the maximum likelihood over ${\vec{\lambda}}$ which serves as the detection statistic. For additive Gaussian noise, the likelihood $\Lambda$ is given by \cite{Finn-1992}
\begin{equation}
  \Lambda \big( \vec{\lambda} \big) = \mathrm{exp}\Big\{\big<s\big|h(\vec{\lambda})\big> - \frac{1}{2}\big<h(\vec{\lambda})\big|h(\vec{\lambda})\big>\Big\}.
  \label{eq:likelihood}
\end{equation}

The inner product $\big< s \big |h \big>$  is the complex cross-correlation \cite{Dhurandhar-1991} between the detector output $s(t)$ and the gravitational wave signal $h(\vec{\lambda})$ weighed inversely by the noise power spectral density of the detector \cite{Babak-2013}:
\begin{equation}
 \big<s \big| h(\vec{\lambda})\big>_{\Delta t} = 4 \; \mathrm{Re} \int_{\flow}^{\fhigh}\frac{\tilde{s}^{*}(f) \; \tilde{h}(f; {\vec{\lambda}})}{S_{n}(f)} e^{-2\pi if\Delta t} \  df,
 \label{eq:fft_conv}
\end{equation}
where $S_{n}(f)$ is the one-sided noise power spectral density defined by $\big< \tilde{n}(f) \big | \tilde{n}^{*}(f^{'})\big> = \frac{1}{2}S_{n}(|f|)\;\delta(f-f^{'})$, asterisk $^*$ denotes the complex conjugation operator, the frequency range $\flow \leq f \leq \fhigh $ marks the effective bandwidth of the detector and $\Delta t$ is the time delay between these two signals. The signal-to-noise ratio (SNR) $\rho$ after filtering $s(t)$  is defined as:
\begin{equation}
 \rho(\Delta t; {\vec{\lambda}}) = \frac{\big<s \big|h(\vec{\lambda})\big>_{\Delta t}}{\sqrt{\big<h(\vec{\lambda})\big|h(\vec{\lambda})\big>_{\Delta t = 0}}}.
\end{equation}
Without any loss of generality, we assume that all template waveforms are normalized such that $\big<h(\vec{\lambda})\big|h(\vec{\lambda})\big>_{\Delta t = 0} = 1$. From Eq.~(\ref{eq:likelihood}) it is clear that this allows us to use the log-likelihood function (or equivalently, the SNR) maximized over the parameters, as the detection statistic.

%Further, any difference between the template and signal parameters will lead to a loss of optimal SNR. 
The log-likelihood function can be maximized over all time lags ($\Delta t$) by using Fast Fourier Transform (FFT) based convolution as shown in Eq.~(\ref{eq:fft_conv}). It can be maximized over other extrinsic parameters analytically \cite{Dhurandhar-1991}. On the other hand, as one cannot maximize the log-likelihood function over intrinsic parameters analytically, a brute force approach is needed. In this case, a discrete grid of points is placed to cover the intrinsic parameter space. One evaluates the log-likelihood surface at each of these points and the maximum is suitably determined. 

The template bank consisting of these discrete set of points on ${\bm{\lambda}}$ is constructed using a control parameter $\Mmin$; commonly known as minimal match in GW literature. This parameter is chosen such that the minimum overlap of an arbitrary vector in the signal manifold (within the deemed parameter space) and at least one template in the bank never drops below this value. The art of template placement lies in maximizing the inter-template separation without violating this constraint with the aim of achieving the smallest bank size. In this way, we can map the template placement to the \textit{sphere covering problem} \cite{Conway-1999, Prix-2007} with spherical cells of radius equal to $\sqrt{1 - \Mmin}$. One wants the smallest number of overlapping templates (i.e. \emph{metric spheres}) to fully cover the space (i.e. \emph{leave no holes}).
The equation of such a spherical cell, centered at $\vec{\lambda} \in \bm{\lambda}$, is given by
\begin{equation}
\big<h(\vec{\lambda}) \big| h(\vec{\lambda} + \Delta \vec{\lambda})\big> = \Mmin.
\end{equation}
From the normalization of waveforms, it is clear that $\Mmin \leq 1$. For high values of this parameter, the LHS can be Taylor series expanded upto leading order in the small parameter $\Delta \vec{\lambda}$:
\begin{equation}
1 - g_{\mu\nu} \Delta \lambda^\mu \Delta \lambda^\nu = \Mmin,
\label{eq:match}
\end{equation}
where $\Delta \lambda^{\mu}$ are the components of the vector $\Delta \vec{\lambda}$ and 
\begin{equation}
g_{\mu\nu} = -\frac{1}{2} \frac{\partial^{2}\big<h(\vec{\lambda}) \big| h(\vec{\lambda} + \Delta \vec{\lambda})\big>}{\partial \Delta \lambda^{\mu} \; \partial \Delta \lambda^{\nu}}\Big|_{\Delta \vec{\lambda} = 0}
\label{eq:metricDef}
\end{equation}
is the metric over the signal manifold which is essentially the Fischer information matrix projected on the intrinsic parameter space and calculated using standard covariance matrix method \cite{Poisson-1995, Owen-1996, Owen-1999}. The match between two nearby points in the parameter space can be calculated easily using the metric. Eq.~(\ref{eq:match}) can be re-arranged as $g_{\mu\nu} \Delta \lambda^\mu \Delta \lambda^\nu = (1 - \Mmin)$,
and identified to be the equation of an ellipsoid in 3D centered at a point $\vec \lambda$. For higher dimensions it represents a hyper-ellipsoid. We shall refer to this as the minimal-match ellipsoid elsewhere in the paper. In GW searches from CBC sources, template banks are usually constructed at $\Mmin \sim 0.97$ \cite{Babak-2013}. This corresponds to a loss in detection rate of $\sim 9\%$ assuming uniform distribution of such sources.  

\subsection{State of the art in template placement}
 \label{sec:stateOfArt}
 
As mentioned earlier, the geometric and stochastic template placement algorithms are two broad class of methods used in searches for gravitational wave signals from compact binary coalescences.  

Previous searches for GW signals from nonspinning compact binaries in initial-LIGO and Virgo \cite{psearch1, psearch2, psearch3} data have used the metric based geometrical hexagonal template placement in two dimensions \cite{Cokelaer-2007, Babak-2006, Owen-1996, Owen-1999, Dhurandhar-1991}. Templates are placed in chirp time coordinates $\{\tau_{0}, \tau_{3}\}$ instead of component masses $\{m_{1}, m_{2}\}$, since the templates are almost uniformly spaced in the former. This process starts by initializing a template point and then finding neighboring points in a $A_2^*$ hexagonal lattice, where the centre of the hexagons represent the position of individual templates. Hexagonal tiling offers the most efficient space filling in two dimensions, which optimizes the number of template points and in turn reduces the total computational cost of the search. To construct this geometric bank one require the semi-analytic metric on the signal manifold assumed to be slowly varying over the parameter space. The curvature effects leads to some loss of efficiency in this strategy. This geometric placement was initially demonstrated \cite{Cokelaer-2007} for 2PN SPA family of waveforms. At present, the metric for 3.5PN SPA waveforms \cite{Keppel-2013}  is available which allows template placement for such waveforms as well.

Geometric template placement in higher dimensional intrinsic parameter space (e.g. component masses and spins) has several problems. First, the metric may not be available for such signal manifolds and secondly, (unlike the hexagonal packing in 2D), optimal geometrical placement in higher dimensions with curvature are not well known. It is further complicated from curvature and boundary effects. To mitigate the curvature issues which leads to rapidly changing metric components over the parameter space, new coordinates are being explored in which the metric is slowly varying \cite{Ajith-2014}, but they do not solve the problem completely.
%As noted above, it is considerably harder to place templates geometrically for precessing spinning searches, where lattice based geometric method cannot be used and where the stochastic method has computational limitations. 
Recent studies \cite{Brown-2012, Harry-2014} have explored geometrical placement in higher dimensions for aligned-spin BNS and NSBH systems for some specific waveform families. In this method, one constructs a metric on the parameterized coordinates - taken to be the coefficients of the 3.5PN TaylorF2 expansion of the orbital phase \cite{Archana-2013}, instead of the usual chirp time coordinates. Since the metric is globally flat in these new coordinates, one can globally transform it into a Euclidean coordinate system. Finally, a principal coordinate analysis facilitates the projection to an effective lower dimensional parameter space, which can be covered by a grid placed in a hexagonal lattice. For NSBH \cite{Harry-2014} systems, the parameter space is covered by stacking several two dimensional hexagonal lattices along the direction of the minor axis. This method is available in the PyCBC software package \cite{pycbc, Usman-2015, Canton-2014, findchirp}. 

An alternative approach of template placement is the so-called stochastic method \cite{Harry-2009} where one starts with an empty template bank (or a set of seed points) to which random points, drawn from a uniform distribution over the deemed parameter space, are appended in an iterative fashion. At every iterative step, a new random point is proposed to be included in the template bank: this proposal is rejected if it happens to lie too close to the points that are already in the existing list otherwise it is accepted. For each accepted proposal, the rejection rate is determined and the process terminates when this rate exceeds a certain threshold averaged over the last few acceptances. We shall call this the \textit{bottom-up} approach in building the list of templates by considering random proposals one by one, retaining only the valid ones. This has been encoded in the LSC Algorithm Library (LAL) software suite via the program \texttt{lalapps\_cbc\_sbank}.

The stochastic bank placement algorithm can also be cast in a top-down fashion. In this alternate implementation, one starts with an empty template list $\tlist$ and a list of very large number of proposals $\rlist$ distributed uniformly over the deemed parameter space. One picks a random point $r \in \rlist$ and appends it to $\tlist$, following which all points from $\rlist$ that lie within the minimal match distance from $r$ are removed. The process continues in an iterative manner until all points from $\rlist$ are exhausted. We call this the top-down approach as the template-bank is microfabricated out of a large block of random proposals by paring it down to the desired shape and order. 

Both these methods are algorithmically equivalent - however, the top-down approach is more useful in projecting a geometric structure over the stochastic template bank leading to the hybrid geometric-random placement algorithm described in this paper. In Table \ref{tab:bankSummary} we demonstrate that the top-down approach is also faster by a factor $\gtrsim 2$ over the traditional bottom-up approach as one is able to eliminate many proposals for a single accepted proposal using efficient computational data structures such as binary search trees (BST \cite{bst1}). The number of templates generated by both these methods is nearly identical. 

The stochastic method is relatively easier to implement and does not require the metric over the signal manifold \textit{per se}. The distance between the proposed point and the elements of the current template bank can be directly calculated by evaluating the match inner product Eq.~(\ref{eq:fft_conv}). This brute-force approach for match calculation allows it to be extended easily to higher dimensional parameter spaces and overcome irregular boundary effects. The disadvantages of stochastic method include the requirement of high computational time as several million proposals have to be processed to guarantee adequate coverage and the fact that it generates substantially more templates than the geometric bank. The computational time can be reduced, if we use the metric (if available) for match calculation. But the intrinsic stochastic nature of the algorithm leads to the inefficiency in grid placement.

Another instance of template placement developed for aligned spin binary black hole (BBH) searches has explored a combination of  geometric and stochastic approaches \cite{Capano-2016, Privertia-2014}. In this method, at first one generates an aligned-spin geometric hexagonal lattice template bank upto some valid range of parameters which is then used as a ``seed" bank for the stochastic placement - thereby accelerating the placement. This method generates $\sim 5.5 \%$ fewer template points than the stochastic method and has been used as part of the uber template bank used in CBC searches in the data from the first observational run of aLIGO \cite{gw15092014:CBC}.
%This has been used in the O1 searches for BBC searches with Advanced LIGO data \cite{gw15092014:CBC}.

\section{A new geometric random algorithm for template placement}
\label{sec:algorithm}

In this section we present the metric based hybrid geometric-random template placement algorithm in three dimensions using a truncated octahedral lattice. Such lattices are the Dirichlet-Voronoi polytope of body-centred cubic $A_3^*$ lattice \cite{Schurmann-2006}. The latter provide optimal coverage for \textit{conformally flat} spaces where the metric-coefficients are constant \cite{Prix-2007}. 
It is interesting to note that this is in line with Lord Kelvin's conjecture \cite{kelvin} according to which, truncated octahedron based space filling is optimal in flat 3D space. 
A truncated octahedron is a 14-sided space filling polyhedron and has the highest volumetric quotient (Eq. \ref{eq:Qv}) which makes it suitable for the lattice structure. Other geometric properties are tabulated in Appendix \ref{appendix:TO}.

Such a template bank would be applicable for gravitational wave searches from the compact binaries described adequately by their component masses and a single ``reduced spin" parameter $\chi_r$ using $\tfrs$ signal model. This signal model is constructed using  post-Newtonian  (PN)  template  family  of  gravitational  waveforms  from  inspiralling compact binaries with non-precessing spins. Here, the spin effects are captured by the parameter $\chi_r$ defined as a mass-weighted, linear combination of individual dimensionless spin magnitudes. The details of the signal model can be looked up in Appendix \ref{appendix:sigmodel}. 

In order to construct the three-dimensional geometric random template bank, we require the metric on the parameter space. The metric  $g_{ij}$ for $\tfrs$ approximant varies rapidly in  $\{m, \eta, \chi_{r}\}$ coordinate system where $m$ denotes the total mass and $\eta$ denotes the symmetric mass ratio. In order to enhance the efficiency of the algorithm, one places the templates in dimensionless chirp time coordinates $\{\theta_{0}, \theta_{3}, \theta_{3s}\}$, in which the metric components are slowly varying over the parameters. These new coordinates are defined as:

\begin{equation}\label{eq:ctf}
 \begin{split}
  \theta_{0}	&= \frac{5}{2^{\frac{1}{3}}} \bigg(\frac{1}{16\pi f_{0}m\eta^{3/5}}\bigg)^{\frac{5}{3}} \\  
  \theta_{3}	&= \bigg(\frac{16\pi^{5}}{25}\frac{\theta_{0}^{2}}{\eta^{3}}\bigg)^{\frac{1}{5}} \\
  \theta_{3s}	&= \frac{113\chi_{r}\theta_{3}}{48\pi}
 \end{split}
\end{equation}

However it is to be noted that the curvature effects do not vanish completely due to the above coordinate transformations. As such, Kelvin's conjecture does not hold directly in this non-flat space. In the following sections, we show that the truncated-octahedral design can still be used for spaces with slowly varying curvature.
This is achieved by merging the stochastic bank placement algorithm along with a local $A_3^*$ lattice.

The method outlined in Algorithm (\ref{algo:gr}) proceeds by initializing  three lists as follows:
\begin{enumerate}
\item[(a)] $\rlist$: a list of uniformly distributed random points sprayed over the deemed parameters space
\item[(b)] $\tlist$: an empty list for template points
\item[(c)] $\klist$: an empty temporary list  
\end{enumerate}

At first $\klist$ is initialized or seeded with a random point chosen from $\rlist$ which is immediately appended to the list of templates $\tlist$ as its first element. 
%Then the points in $\rlist$ that are too close to the first element of $\klist$ are pruned. 
The possible 14 TO neighbours of this initial point are then calculated and appended to $\klist$ followed by the removal of the first element. Only those TO neighbors which fall within the parameter space and are farther than $\mathcal{M}_{min}$ from existing points in the $\tlist$ and $\klist$ can be appended. This completes one iteration of the inner loop of the algorithm. We continue these steps until $\klist$ becomes empty. The latter happens when no new valid TO neighbors are appended in successive iterations of the inner loop of the algorithm which ultimately exhausts all the points in $\klist$. 
At this stage, all points from $\rlist$ that are within the minimal match distance from the elements in $\tlist$ are deleted; following which $\klist$ is re-seeded with a new random point from $\rlist$. 
Termination of template placement algorithm occurs when $\rlist$ is exhausted. 

\algrenewcommand\algorithmicindent{2.5em} % Change indentation size
\begin{algorithm}[H]
\caption{Geometric-Random Template Placement}
\label{algo:gr}
\begin{algorithmic}[1]  % [1] is used to show line numbers
\STATE generate $\rlist$
\STATE $\klist = [\ ]$
\STATE $\tlist = [\ ]$
\WHILE {$(\rlist)$}
    \STATE $\klist \leftarrow$ random point $ p \in \rlist$
    \WHILE {$\klist$}
	      \STATE $\tlist \leftarrow$ append $\klist [0]$
	      \STATE find all possible TO neighbours of $\klist [0]$
	      \STATE $\klist \leftarrow $ append {\emph{valid}} TO neighbours
	      \STATE delete $\klist [0]$
	\ENDWHILE
    \STATE delete all minimal match neighbours of $\tlist$ from $\rlist$
\ENDWHILE
\end{algorithmic}
\end{algorithm} 

The advantage of casting the stochastic template placement algorithm (see Sec \ref{sec:stateOfArt}) in a top-down fashion is clear when we consider the extreme case where no TO lattice neighbours can be found for any proposal point in $\rlist$. In this case, it is clear from Algorithm \ref{algo:gr} that the geometric-random algorithm naturally falls back to the vanilla-stochastic placement algorithm. This brings in the robustness of the latter to the proposed new algorithm.  This contingency arises for a small fraction ($\sim 5-10 \%$) of points in $\rlist$ - due to boundary edge effects and small uncovered patches arising from curvature effects. By construction the geometric-random placement method presented here is more efficient than vanilla-stochastic algorithm.

Also note that the TO, whose neighbors are added to $\klist$ in Lines 8-9 of Algorithm (\ref{algo:gr}) refers to the one inscribed inside the minimal match ellipsoid given in Eq.~(\ref{eq:match}) above. The gradual change in curvature (changing metric components) leads to changing orientation and size of these ellipsoids which affects the size and  orientation of the inscribed truncated-octahedron. This in turn, affects the coordinates of its 14 neighbors. In this way the placement algorithm automatically responds to the curvature effects. This is equivalent to assuming flat local patches of the signal manifold which can be optimally covered by $A_3^*$ lattice. 

\subsection{Implementation}
\label{sec:imp}
We now highlight some salient features for efficient implementation of the algorithm.

\subsubsection{Initializing \texorpdfstring{$\rlist$}{} with uniform random points}

In order to generate a random list of points over the parameter space, at first we calculate the minimum and maximum possible values of dimensionless chirp times $\theta_{0}, \theta_{3}$ and $\theta_{3s}$ for the range of parameters over which the bank is to be placed. 
A set of random points are generated in $\{m_{1}-m_{2}\}$ space along the boundaries marking the constraints for component masses, chirp masses and mass ratios. Afterwards, a coordinate transformation is taken to the dimensionless chirp time coordinate system $\{\theta_{0}-\theta_{3} \}$ which allows the calculation of the extremities in these coordinates. The extreme values of $\theta_{3s}$ are evaluated using $|\chi_{r}| \leq 1$ in Eq.~(\ref{eq:ctf}). 
Once these extremities are known, uniform random points are proposed within these values - out of which only those which lie within the specified mass and spin ranges are retained. 

The final banksize depends strongly on the initial density of points in $\rlist$ upto a critical value, beyond which it does not change very much. Empirically, we find this number to be about $\sim 1/10$ of the total volume calculated in the dimensionless coordinates. The latter depends on the metric and varies across the parameter space due to which we estimate an average cell volume by sampling different points across this space. The total number of cells needed to cover the space is given by the ratio of the total volume divided by the average volume of a cell. Monte Carlo integration method can be used to estimate the total volume. 

\subsubsection{Deletion of points within minimal-match ellipsoid}
\label{sec:pointsDel}

We discuss an efficient way to delete points from $\rlist$  that lie within minimal match distance from the template bank points.  
For a fixed minimal match value, Eq.~(\ref{eq:match}) represents the surface of an ellipsoid with semi axes length:
\begin{equation}
 A_{i} = \sqrt{\frac{1-\Mmin}{e_{i}}},
\end{equation}
where $e_{i}$'s are eigenvalues of the metric $g_{ij}$ evaluated at the ellipsoid centre. 
The orientation of the ellipsoid is defined by matrix $\mathcal{R}$ composed of eigenvectors of the metric $g_{ij}$, where $\mathcal{R}_{ij}$ is $j^{th}$ component of the $i^{th}$ eigenvector. 
To visualize the   deletion process, let us consider a cell centred around a template. The points in $\rlist$ lying inside the cell can be determined by calculating the match between this template with the all other points in $\rlist$ but such a brute force approach will not be computationally efficient.
We need to find a smaller set of points against which we can identify these neighbors more efficiently.  To this end, we can first use binary search tree (BST) \cite{bst1, bst2, bst3} data structure to identify a smaller list of points from $\rlist$ that lie within a sphere of radius equal to the largest semi axes of the cell. From this smaller list of points, we then proceed to delete the ones that satisfy $g_{ij}\Delta\lambda^{i}\Delta\lambda^{j} \leq 1 - \Mmin$.

This can be further refined by binning the template points in $\theta_0$ such that each bin contains about $\sim 1000$ templates. For each such subset, the removal of points from $\rlist$ proceeds by identifying those points that lie in the same bin (within some acceptable margin in $\theta_0$) and then applying the above strategy to eliminate the points. This refinement is made possible due to the fact that match values decrease monotonically with the difference in parameters and is most sensitive to changes in $\theta_0$. Additional binning in other two coordinates may possibly improve the computational efficiency of removing the points from $\rlist$ even more.

\subsubsection{Global co-ordinate transformation}
\label{sec:gtf}

In dimensionless chirp time coordinates, most of the ellipsoidal cells have semi-axes ratio around $1000:10:1$. This implies that the number of points inside this flat and elongated ellipsoid is a small fraction of the total number of points contained in the sphere with radius equal to semi-major axis. This undermines the efficiency of deleting the points using the technique outlined in Section \ref{sec:pointsDel}. We use a conformal coordinate transformation in such a way that one of the cells transforms to a unit sphere. Note that due to curvature effects, the same transformation when applied to other cells does not guarantee them to change to unit spheres but mitigates the problem of highly asymmetric semi-axes ratio to a large extent.  We use the eigenvalues and eigenvectors of the metric at a fiducial point to transform the coordinates ($\theta_k \rightarrow \xi_k$) as,
 \begin{equation}
\xi_{i} = \underset{j,k=1}{\sum^3}\mathcal{S}_{ij}\mathcal{R}_{jk}^{T}\theta_{k} 
\label{eq:globalTrans}
 \end{equation}
where, $\mathcal{S}_{ij} = \sqrt{e_{i}}\delta_{ij}$ is a diagonal scaling matrix and the rotational matrix $\mathcal{R}_{ij}$ is constructed from the $i^{\text{th}}$ component of the $j^{\text{th}}$ eigenvector of the metric calculated at the fiducial point. 
This is carried out by first binning the templates in $\theta_0$. The transformation in 
Eq. (\ref{eq:globalTrans}) is carried out using the metric of the centre point in each bin.

The metric at all other points transform as:
\begin{equation}
 \bar{g}_{ij} = \underset{p,q,r,t=1}{\sum^3}\mathcal{S}_{ip}^{-1}\mathcal{R}_{pq}^{T}g_{qr}\mathcal{R}_{rt}\mathcal{S}^{-1}_{tj}.
\end{equation}
It is trivial to check that the metric at the fiducial points transforms to a $3\times 3$ identity matrix.

\begin{figure*}
    \centering
    
    \subfloat[Before global transformation]{\label{fig:gtf_left}%
    			{\includegraphics[width=0.45\textwidth]{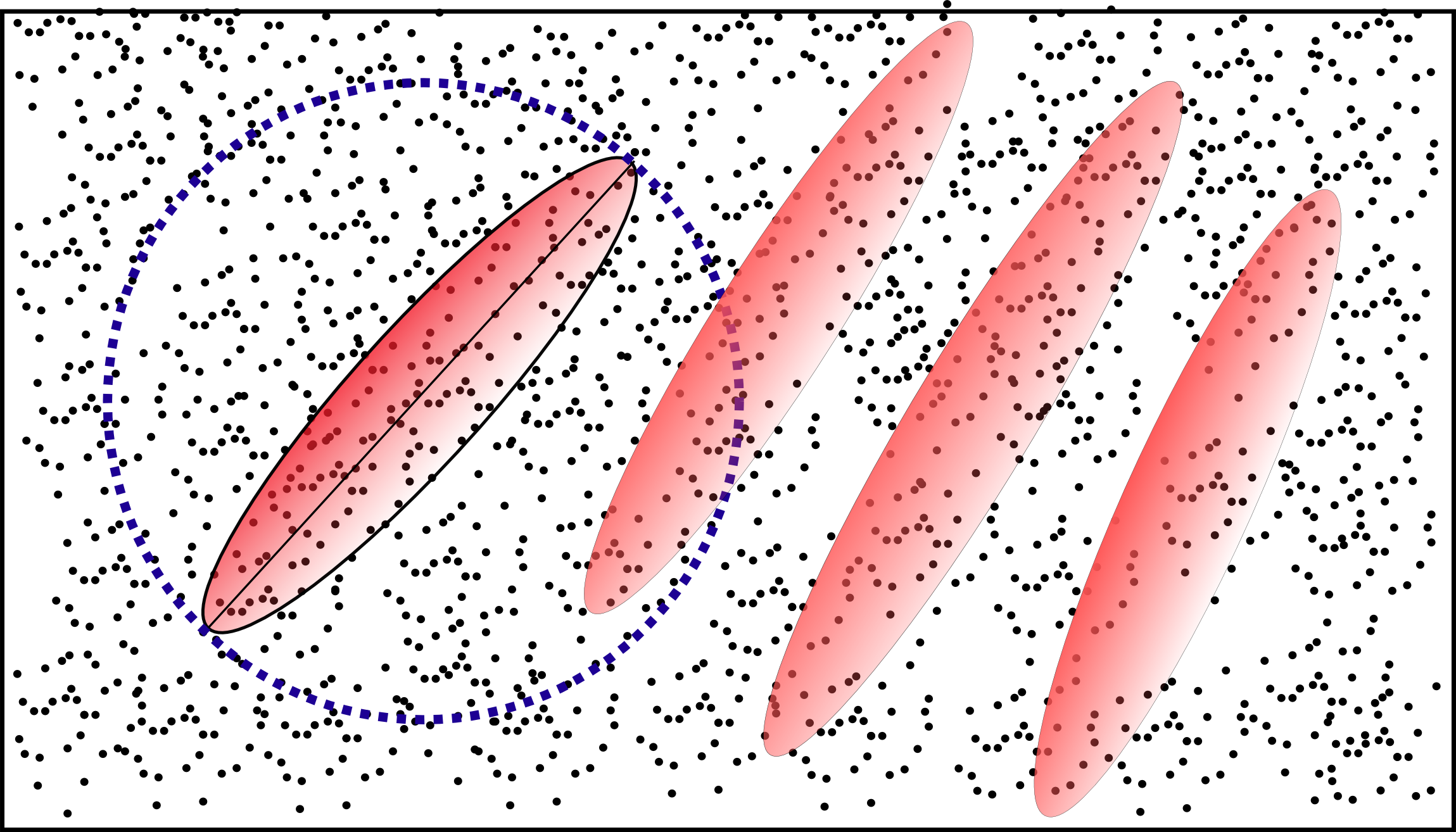} }}%
    \quad
    \subfloat[After global transformation]{\label{fig:gtf_right}%
    			{\includegraphics[width=0.45\textwidth]{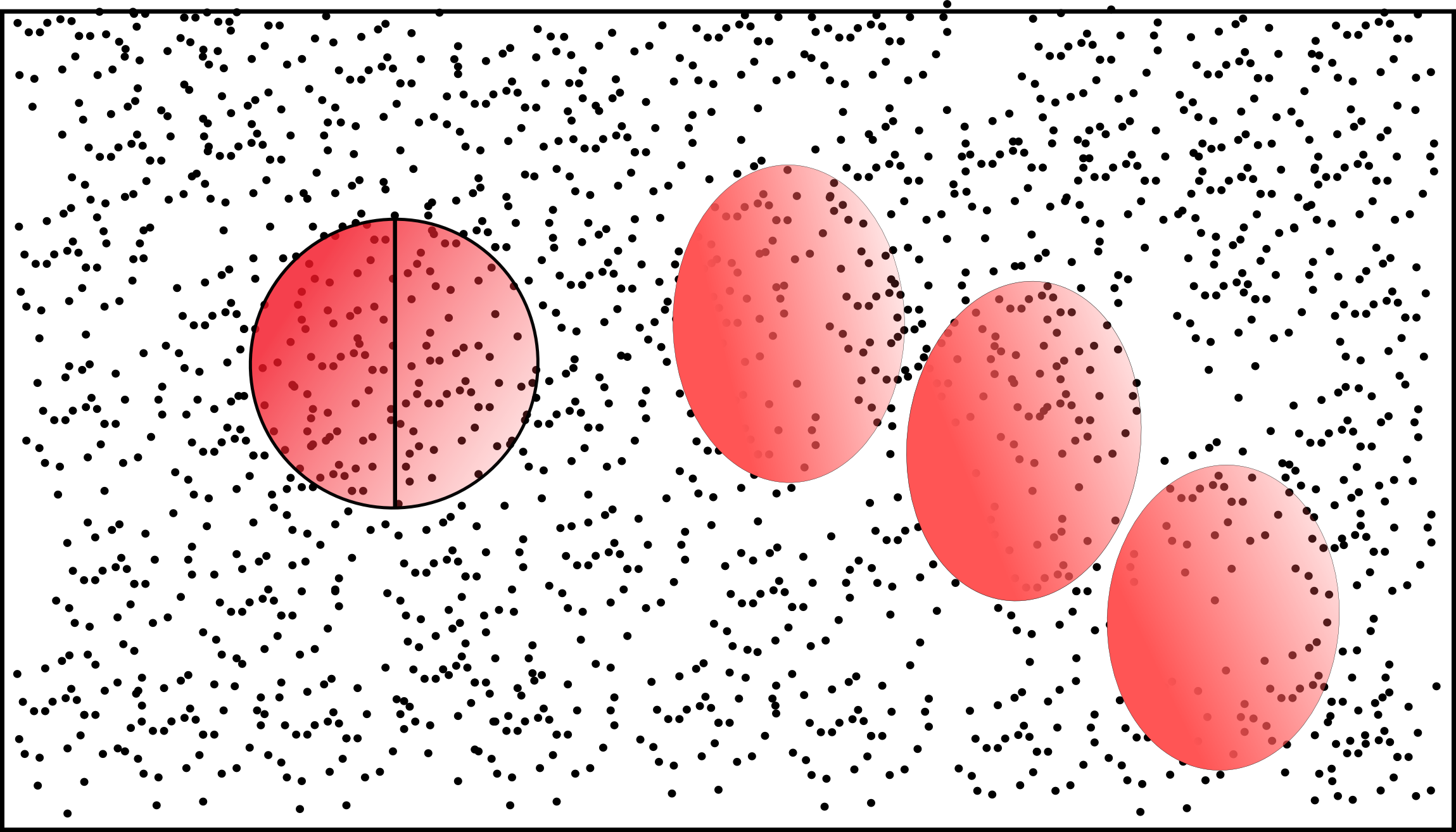}}}%
    					   
    \caption{An example of global transformation Eq.~\ref{eq:globalTrans}) to speed up the point elimination part using BST algorithm. Here we construct the global transformer using the metric of an ellipse which is inscribed in blue dashed circle of the left figure as shown in the left panel. After the transformation the same cell becomes a unit sphere and nearby cells also become more spherical.}
    \label{fig:gtf}%
\end{figure*}

This global coordinate transformation Eq.~(\ref{eq:globalTrans}) causes other nearby cells to become almost spherical  with semi-axes in the ratio  $\sim \ 1:1:1$. In Fig.~\ref{fig:gtf} we show the effect of global coordinate transformation on nearby cells. This is doubly advantageous: not only does the ball-point query volume for the BST searches decrease (leading to fewer points to deal with); the points that lie within the inscribed minimal match ellipsoid now occupy a large fraction of its volume leading to increased efficiency in removal of points.

\subsubsection{Locally placed truncated octahedral lattice}

The template placement problem in 3D can be mapped to the tiling problem such that a minimum number of similar cells is used without leaving any region uncovered. As such, truncated octahedral lattice becomes a natural choice for this problem. 

\begin{figure}
\centering
  \includegraphics[width=0.50\textwidth]{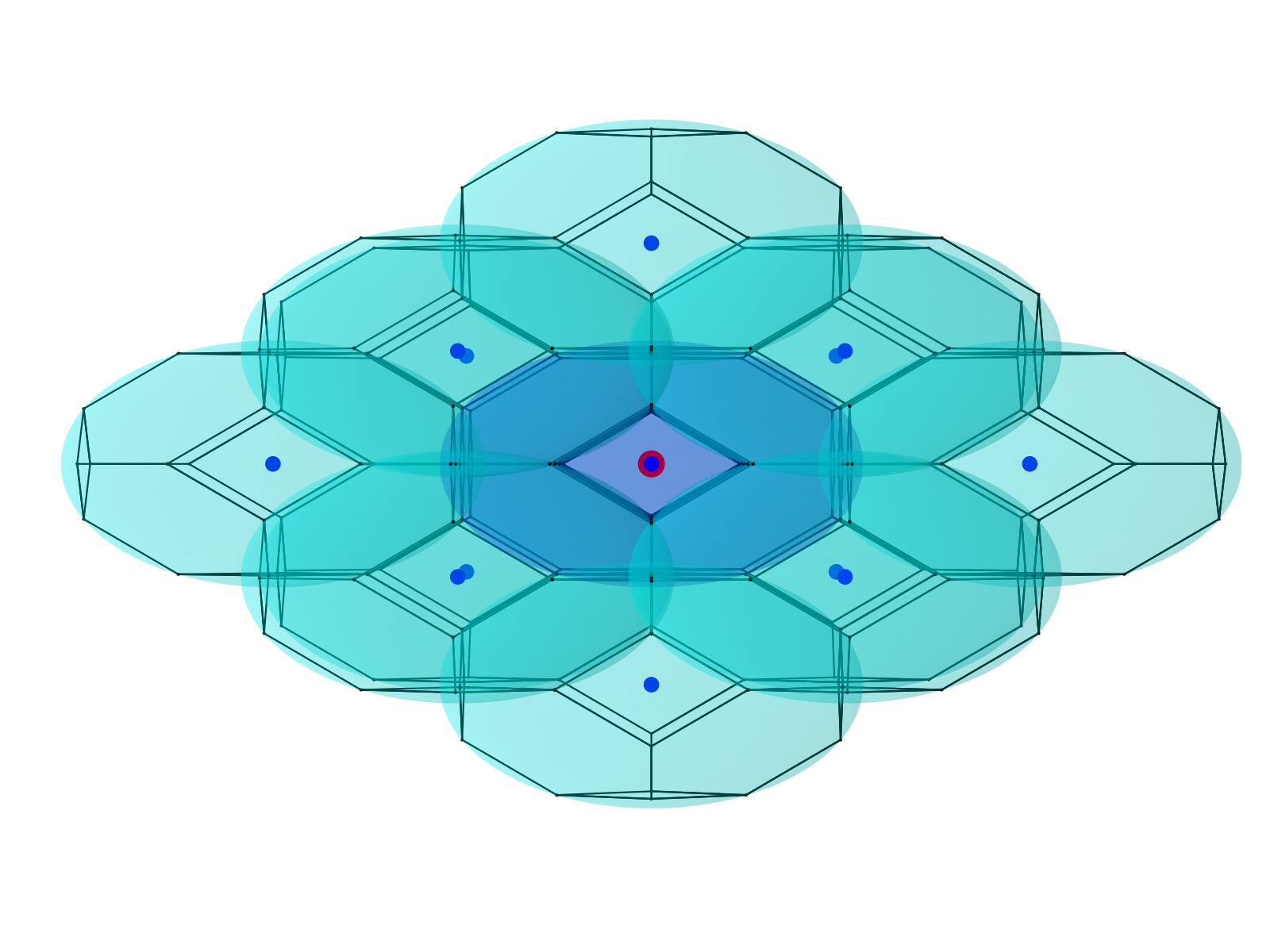}
  \caption{Lattice neighbors of a truncated octahedron inscribed in a elliptical cell. Here only 9 neighbours are visible and the remaining 5 are on the opposite side.}
  \label{fig:elpngb}
\end{figure}

As stated earlier, the TO shares 14 faces with its neighbours in the lattice. Assuming a TO inscribed in a sphere of radius $R$, the coordinates of its 14 lattice neighbors $N_k^{p}$ are available in Table \ref{tab:tongb}. Here $k=(1,2,3)$ is an index on the coordinates of the $p-th$ neighbor. When mapped to the template placement problem, we need to consider TO's inscribed within the minimal match ellipsoids. In this case, the coordinates of the lattice neighbors $\bar{N}_i^{p}$ can be calculated using the rotation and scaling matrices:
\begin{equation}
 \bar{N}_{i}^{p} = \sum_{j,k=1}^3 \mathcal{R}_{ij}^{T}\mathcal{S}_{jk}N_{k}^{p}
\end{equation}
Fig.~\ref{fig:elpngb} shows the 14 neighbours of a TO inscribed in a elliptical cell.   

Due to boundary effects, all 14 TO neighbors of a point need not necessarily be part of the template bank. The following conditions must be checked for:
\begin{enumerate}
  \item[(a)] The point is inside the deemed parameter space and also satisfies $\eta \leq 1/4$. The latter corresponds to the condition that $\theta_k$'s can be inverted to yield physical masses.
\item[(b)] The point is not within the minimal match distance of existing points in $\tlist$ and $\klist$. 
\end{enumerate}
Check (b) above ensures that we do not double count the neighbors. 

As shown in Algorithm (\ref{algo:gr}), we start from a random point in the parameter space (by seeding $\klist$) and tessellate with local TO lattice. Because of curvature and boundary effects, it is not guaranteed that these tessellations cover the entire parameter space. This is marked by the exhaustion of $\klist$ as the placement proceeds. At this stage, we need to re-seed $\klist$ and continue the process iteratively until all points in $\rlist$ have been used up.

\subsubsection{Choice of initial point and variations in banksize}

The template placement algorithm starts from a randomly chosen point in the parameter space by seeding $\klist$ which is copied over as the first element of the template bank list $\tlist$. 

One can start from any point in parameter space: in the present work we have started from ``mid-point" corresponding to component masses $m_{1,2} = (m_{1,2}^{max} + m_{1,2}^{min})/2$ with individual spins $\chi_{1,2} = (\chi_{1,2}^{max} + \chi_{1,2}^{min})/2$. We have checked that starting from other points (e.g. the extremities of the parameter space) also work quite well. This initial choice results in a minor fluctuation of the template bank size and it is recommended that we start from a point well inside the deemed parameter space where the local variations in curvature are less. This ensures maximum tiling before we need to re-seed $\klist$. 

We demonstrate these fluctuations in template bank size by constructing several template banks for compact binary systems whose bank parameters are given in Set II of Table \ref{tab:tmp} starting from different locations in the deemed parameter space. The random seed used to initialize $\rlist$ was kept the same to eliminate bias. The five different starting points were taken to be the centre and extremities of the parameters space respectively and as expected, different choices of the initial seed point resulted in slightly different number of templates in the final bank. We generated an average of $111,257$ templates with $\sim 3\%$ fluctuation which is quite insignificant.

\section{Test of the Algorithm}
\label{sec:testOfAlgorithm}

In this section we construct a template bank using our algorithm and demonstrate its validity for CBC searches. We also compare its performance against a bank generated using the vanilla stochastic method available in the LSC Algorithm Library Suite (LALSuite) \cite{lalsuite}. We need a metric over the dimensionless chirp coordinates and at present two such models are available for use, namely; $\tfrs$ and IMRPhenomB signal model \cite{Ajith-2015}. Our demonstration makes use of the $\tfrs$ model.

\subsection{Construction of the template banks for reduced-spin binary system}
\label{subsec:bankConstruction}

We generate a template bank using $\tfrs$ metric in $(\theta_{0}, \theta_{3}, \theta_{3s})$ parameter space for aligned spin compact binary system. The range of parameters is chosen such that the mass of the first object lies between $1 - 20 \msun$ with dimensionless spin magnitude in the range $\pm 0.98$.  The mass of the second object is taken between $1 - 3 \msun$ with dimensionless spin magnitude in the range $\pm 0.4$. 
%The range of spin magnitude of the lighter mass covers not only observational $(\sim 0.04)$ but also theoretical $(\sim 0.4)$ spin limits of neutron stars. 
The NS-BH boundary mass condition is satisfied i.e any object with mass $\leq 3\msun$ is considered to be a neutron star with corresponding limit on spin magnitude. This template bank can be used for binary neutron star (BNS) and neutron-star-blackhole (NSBH) searches. 

We construct several template banks by varying the number of uniformly sprayed random points over the parameter space, $\rlist$. We consder sizes of $\rlist$ varying between $1 - 8\times 10^{7}$ points. In all the cases, the placement proceeds from a point corresponding to individual masses $m_{1, 2} = (\frac{21}{2}, \frac{4}{2}) \msun$ and individual spin magnitudes $\chi_1 = \chi_2 = 0.0$ . The full specification of input arguments for template generation are  given in Set-I of Table \ref{tab:tmp}. 

\begin{figure}%
    \centering
    \subfloat[Projection along $\theta_{3s}$]{\label{fig7_a}%
    					{\includegraphics[width=0.485\textwidth]{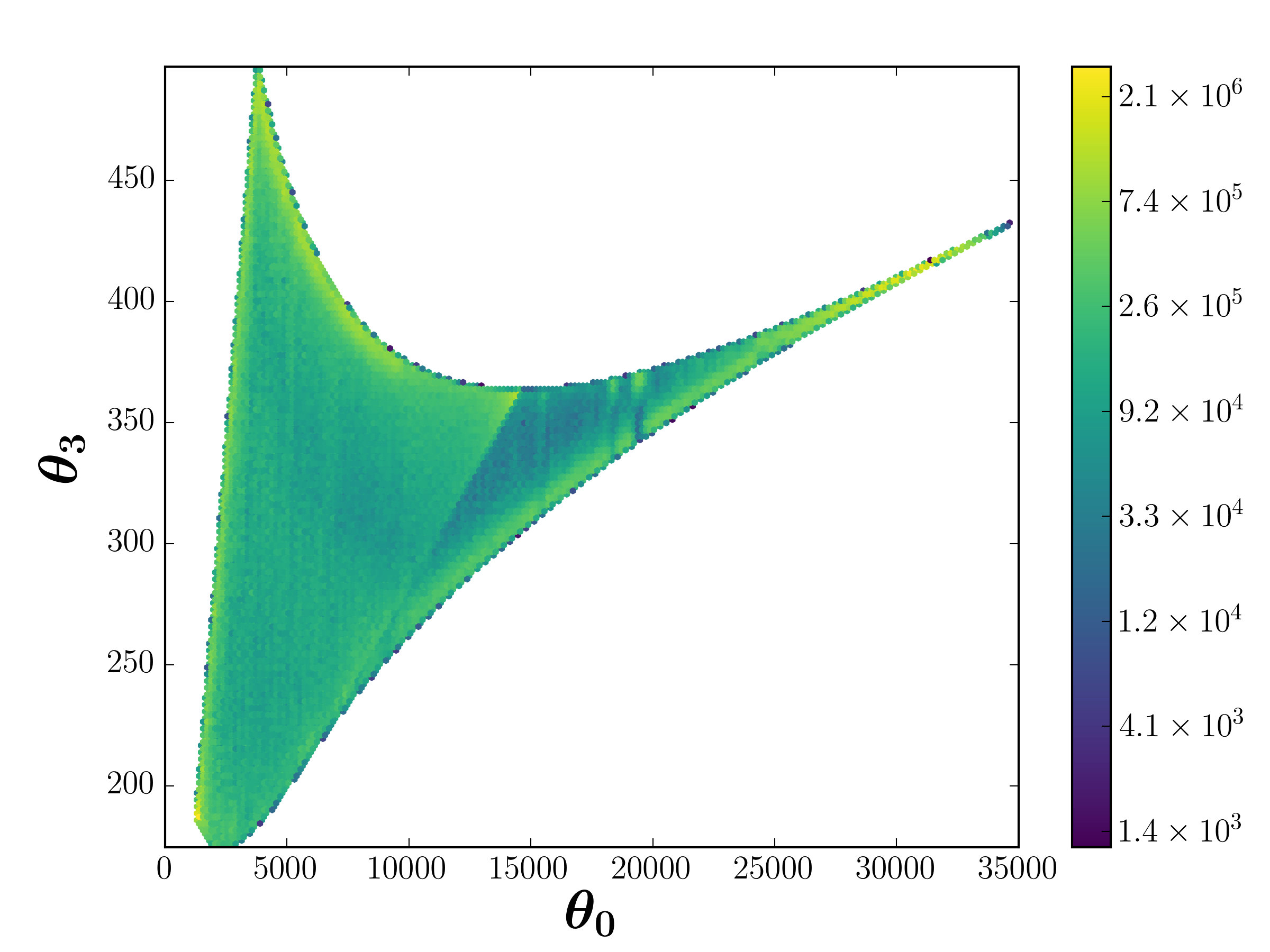} }}%
    \quad
    \subfloat[Projection along $\theta_3$]{\label{fig7_b}%
    					{\includegraphics[width=0.485\textwidth]{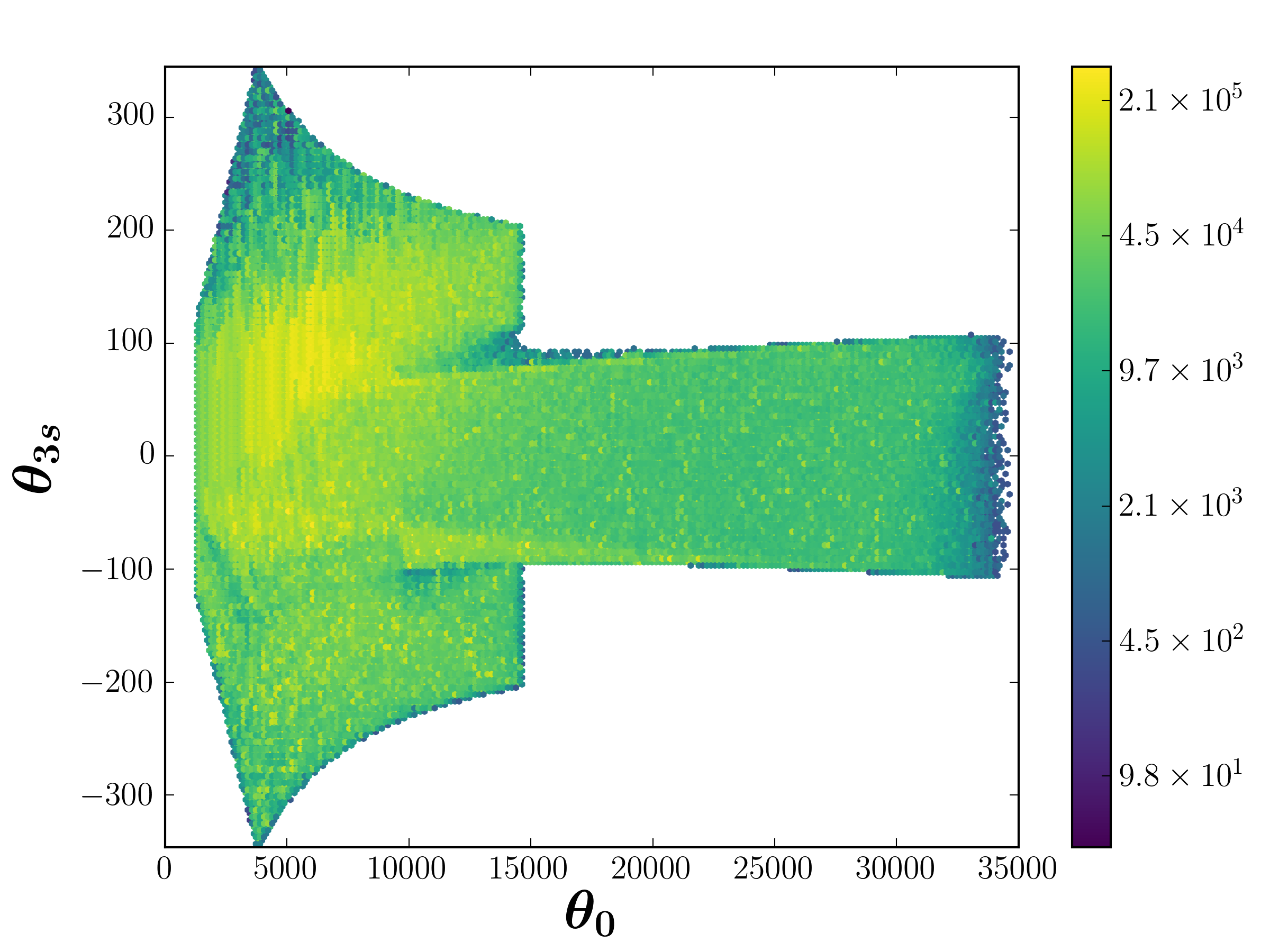}}}%
    \quad
    \subfloat[Projection along $\theta_0$]{\label{fig7_c}%
    					{\includegraphics[width=0.485\textwidth]{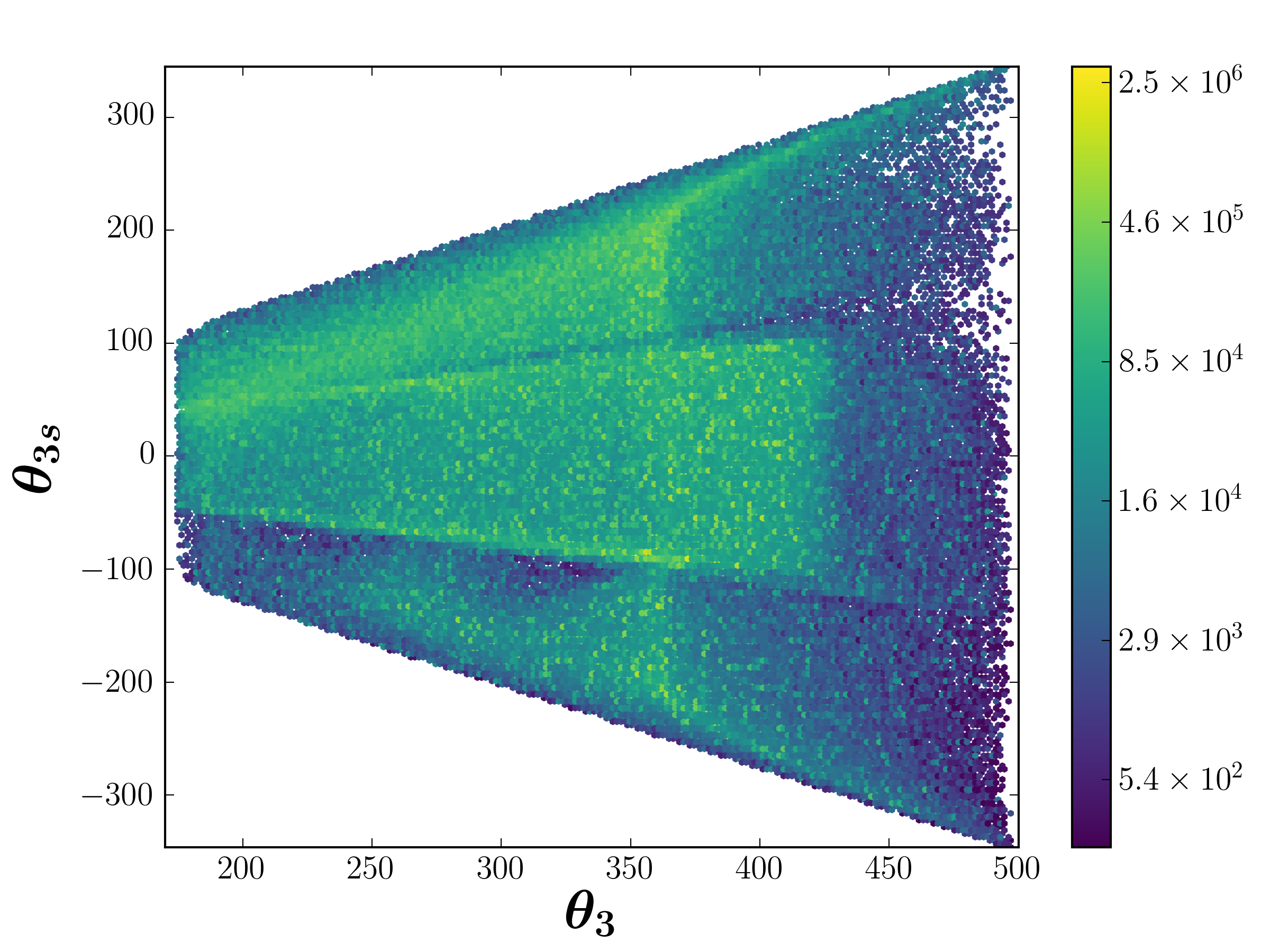}}}%
    					   
    \caption{Area normalized histograms of the template density in various planes. The template bank was constructed using hybrid geometric random algorithm presented in this paper. Each bin of the histogram was normalized by the square-root of the determinant of the metric $\sqrt{|g_{ij}|}$ to ensure equal area. The metric was calculated at the bin centre. The boundary effects are clearly seen. We can also see that bank is highly elongated along $\hat{\theta}_{0}$ direction as compared to both $\hat{\theta}_{3}$ and $\hat{\theta}_{3s}$ directions.}
    \label{fig:tmp}%
\end{figure}

The corresponding final template bank size are listed in Table~\ref{tab:bankSummary}. In Fig.~\ref{fig:tmp} we show the two dimensional projections of such a bank along $\hat \theta_{3s}, \hat \theta_3$ and $\hat \theta_{0}$ directions respectively. The hat $(\hat)$ over the symbols denote unit normal vector along the specified direction. For comparison, we also generated a template bank using the vanilla stochastic method for the same parameters as listed in Set-I of Table \ref{tab:tmp}. The vanilla stochastic bank placement was terminated at a point when there were 1000 rejected proposals for each accepted proposal averaged over the last 10 acceptances. This stochastic bank was found to contain $939,787$ templates which is  $\sim 25 \%$ larger than the geometric-random bank. The computational run time was also recorded and we observed that the geometric-random method took $482$ minutes while the vanilla stochastic method took $3666$ minutes to execute on a single, unloaded processor, \textit{which is $\sim 8$ times faster}. 

\begin{table*}[ht]
  \begin{tabular}{ p{5.3cm} p{4.5cm} p{4.5cm} }
 \hline \hline
 Bank Parameter  						&  Set-I               & Set-II \\
\hline
  Waveform model   						& $\tfrs$              & $\tfrs$ \\
  Noise model 							& $\aligo$             & $\aligo$ \\ 
  Lower cut-off frequency $f_{low}$ 	& $20 \Hz$             & $30 \Hz$ \\  
  Higher cut-off frequency $f_{high}$ 	& $2048 \Hz$           & $1024 \Hz$ \\
  Mass of first object $m_{1}$ 		& $[1,20]\msun$        & $[3,15]\msun$ \\  
  Mass of second object $m_{2}$ 		& $[1,3] \msun$        & $[1,3]\msun$ \\ 
  Spin of first object $\chi_{1}$ 		& $[-0.98, 0.98]$      & $[-0.6, 0.6]$ \\ 
  Spin of second object $\chi_{2}$ 	& $[-0.4, 0.4]$        & $[-0.05, 0.05]$ \\ 
  Size of $\rlist$ 						& $[1-8]\times10^{7}$  & $1\times10^{7}$ \\
  Minimal Match $\Mmin$ 				& 0.97                 & 0.97 \\
 \hline \hline
\end{tabular}
\caption{Parameters used to generate the geometric-random and vanilla stochastic banks. The results for different sizes of $\rlist$ are summarized in in Fig. \ref{fig:fracinj}. In Set-I, the parameter space is chosen by satisfying the NS-BH boundary mass i.e. components with individual mass  $\leq 3\msun$ are identified as NS
with dimensionless spin magnitude in the range $\pm 0.4$.}
\label{tab:tmp}
\end{table*}

\subsection{Validation of the template banks}

We investigate the performance of both the geometric-random and the vanilla stochastic template banks against a set of signal injections from the reduced spin $\tfrs$ signal model. In this section we summarize the results of this comparison and demonstrate that the two banks are nearly identical in performance.

Following Apostolatos \cite{Apostolatos-1995}, the ``fitting factor" $\mathcal{FF}(h_a)$ is defined as a measure of the maximum match over the template bank $\tlist$ for a putative injected signal $h_a$:
\begin{equation}\label{eq:ff}
  \mathcal{FF}(h_{a}) = \underset{h\in \tlist}{\mathrm{max}} \text{\hspace{1mm}}  \big<h_{a}\big| h\big>
\end{equation}

The parameters of the injected signal (chosen from within the deemed parameter space over which the bank is placed) is chosen at random and may not coincide with that of a template point. The mismatch $(1 - \mathcal{FF}(h_{a}))$ indicates the fractional loss of optimal SNR. Because differences between true waveforms and waveform models will always reduce the fitting factor, banks are usually tested to achieve fitting factors slightly larger than the minimal match used in their construction.
The minimal match at which the template banks are constructed ($0.97$ in this case) is somewhat arbitrary - and is arrived by carefully balancing the computational cost of the search against the desired detection efficiency. 
  
\begin{figure}
\centering
    \subfloat[Comparison of hybrid bank performance]{\label{fig4_a}%
    {\includegraphics[width=0.495\textwidth]{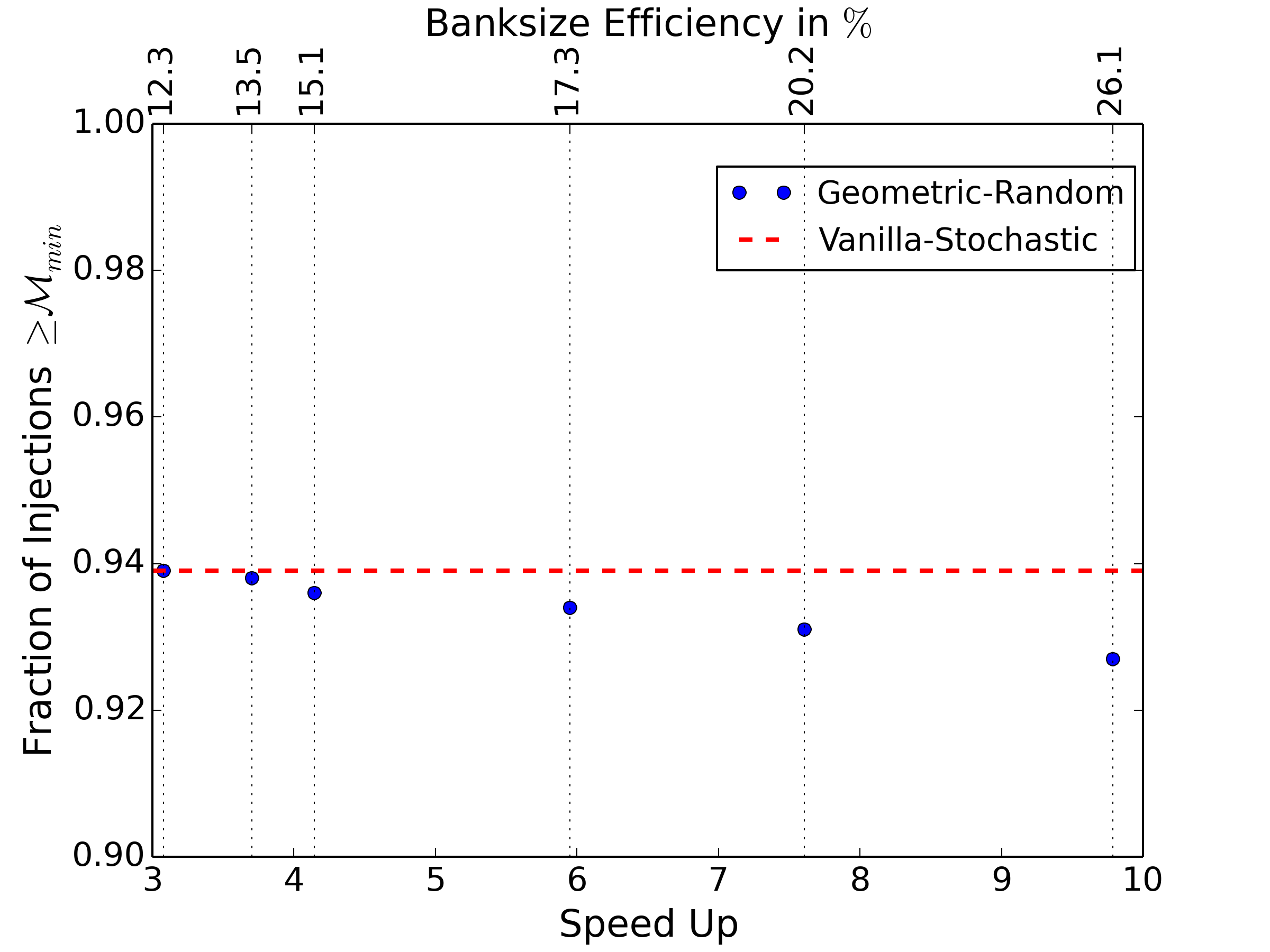}} }%
    \quad
    \subfloat[Fitting factor distribution for hybrid bank]{\label{fig4_b}%
    {\includegraphics[width=0.475\textwidth]{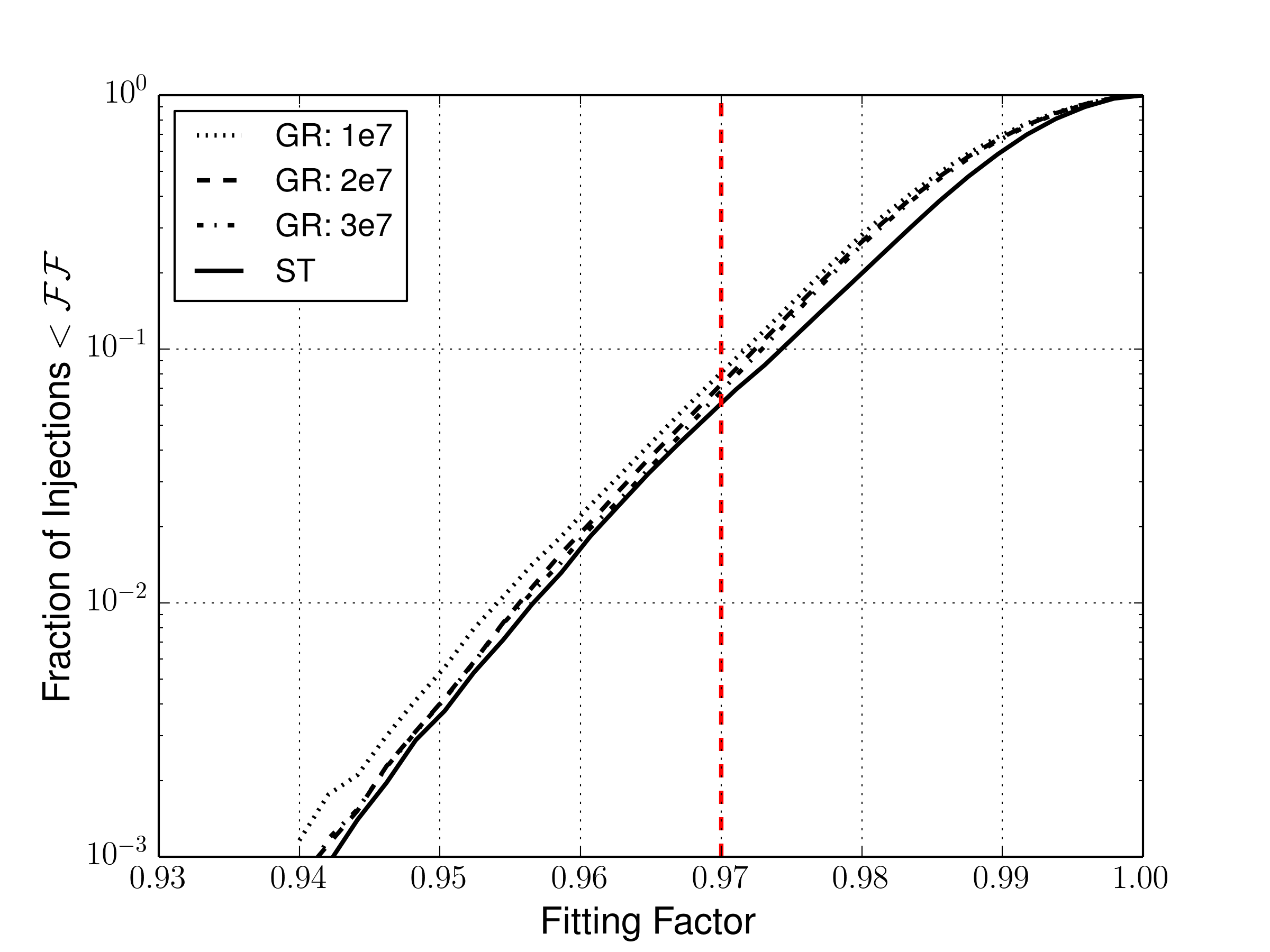}} }%
    \quad
    \subfloat[Fitting factor distribution for hybrid bank]{\label{fig4_c}%
    {\includegraphics[width=0.475\textwidth]{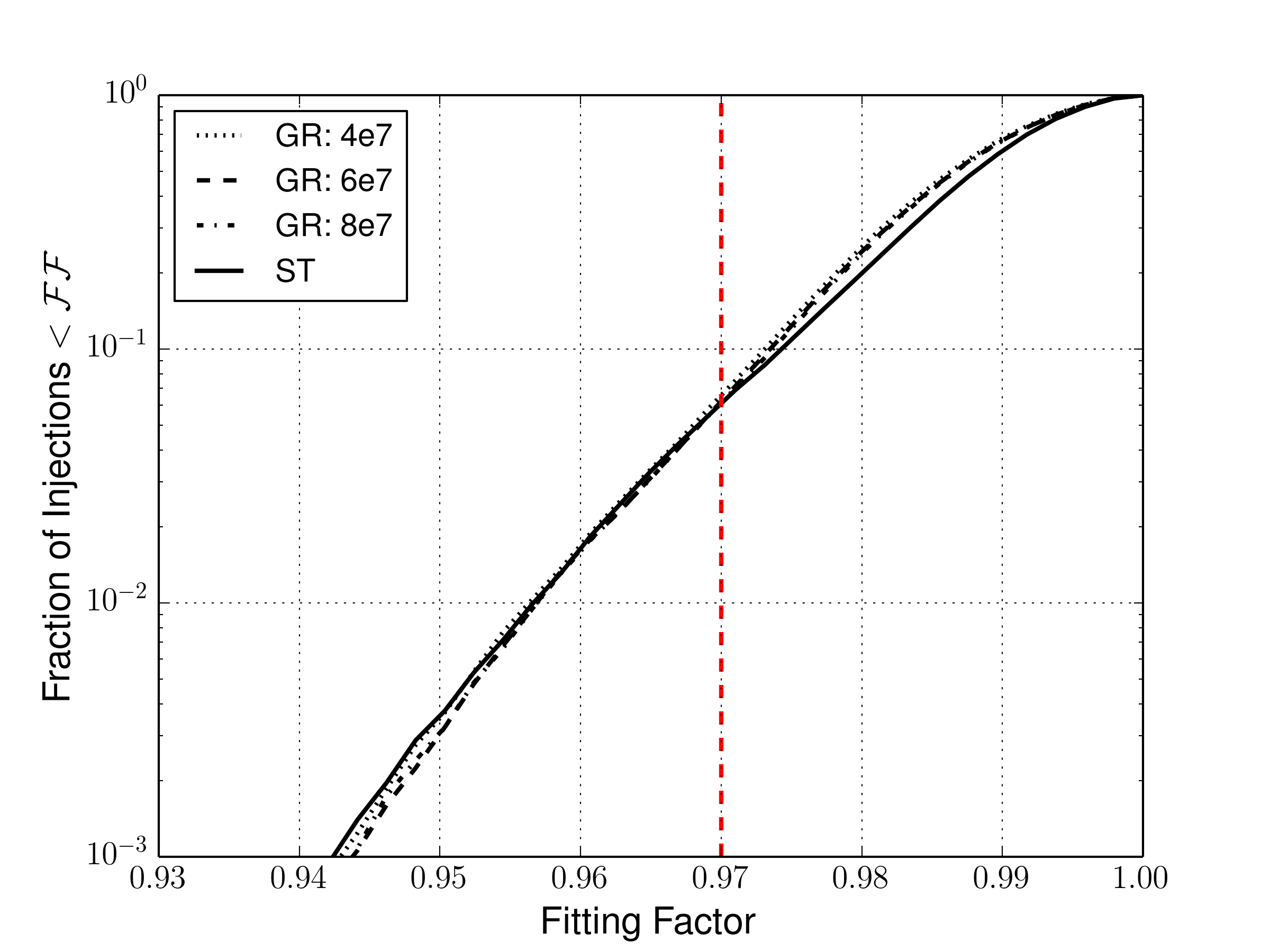}} }%
  \caption{Panel \ref{fig4_a} shows a comparison of the geometric-random bank and the vanilla-stochastic bank constructed over identical parameter ranges is made by plotting fitting factors for $50,000$ $\tfrs$ signals. The horizontal line depicts the percentage of such injections for which the fitting factor is above the bank minimal match in the case of vanilla stochastic bank. The solid dots correspond to the percentage for geometric-random banks constructed with different sizes of $\rlist$ as tabulated in Table~\ref{tab:bankSummary}. The bottom horizontal axis measures the computational speed-up of the hybrid geomtric-random placement while the one on top shows the corresponding efficiency in bank-size. Panel \ref{fig4_b} and \ref{fig4_c} show the cumulative distribution of the hybrid bank fitting factors for the different initial size of $\rlist$. They have been split into two panel for clarity.}
  \label{fig:fracinj}
\end{figure}

For the banks generated as per the parameters in Set-I of Table \ref{tab:tmp}, we injected $50,000$ signals from reduced-spin $\tfrs$ waveform model and calculated the fitting factor using Eq.~(\ref{eq:ff}). The \texttt{lalapps\_cbc\_sbank\_sim} program as implemented in \texttt{LALApps} package of the LSC Algorithm Library Suite (LALSuite) \cite{lalsuite} was used for this calculation. The range of mass and spin parameters of the injected signals were chosen to be the same as that of the bank and were drawn from a uniform distribution in this range. Other extrinsic parameter  were drawn from (i) a uniform random distribution over all possible sky locations (ii) a fixed inclination angle corresponding to edge-on orientation of the plane of the binary, and (iii) a fixed luminosity distance of 1 Mpc. For the banks (corresponding to different initial sizes of $\rlist$), we found that $\sim  92-94\%$ signals were recovered with fitting factor $\geq 0.97$ for both geometric-random bank as well as for the stochastic bank. The results are shown in Fig.~\ref{fig:fracinj}. The fluctuation at higher significant digit was ignored due to the limited sample size of injections. These results are graphically depicted in Fig.~\ref{fig:fracinj} and demonstrate the equivalence of the two banks along with the computational speed up and corresponding efficiency of template placement using the new method.

Although the injection signal model is identical to the one used to construct the template banks, we still find that the fitting factors fall below the minimal match of the bank for a small fraction of the injections. This alludes to the fact that both template banks have \textit{holes} in certain regions of the deemed parameter space where neighbouring templates do not provide adequate coverage. This arises from curvature effects.

The fitting factor depends on the parameters of the injected signals - e.g. masses, spin components, sky position and inclination angle of the binary's plane to the line of sight etc. 
For non-precessing signals including only the dominant radiation mode, the fitting factor
depends only on the intrinsic parameters. To understand the systematics, it is convenient to represent it as a function of two parameters while  averaging out over the remaining ones.

\begin{figure}
    \centering
    
    \subfloat[over component masses]{\label{fig_ff_a}%
    					{\includegraphics[width=0.4825\textwidth]{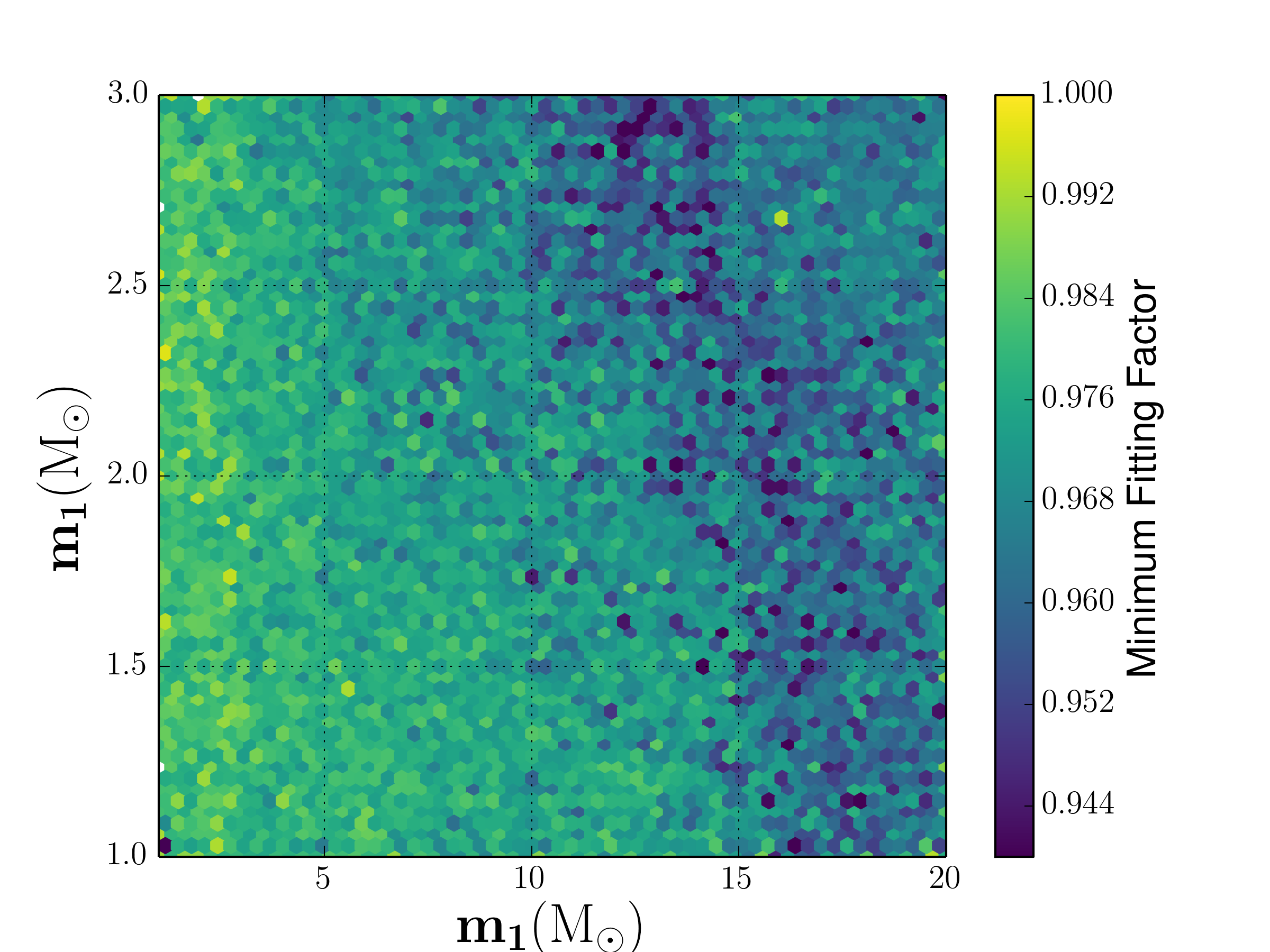} }}%
    \quad
    \subfloat[over total mass and reduced spin parameters]{\label{fig_ff_b}%
    					{\includegraphics[width=0.4825\textwidth]{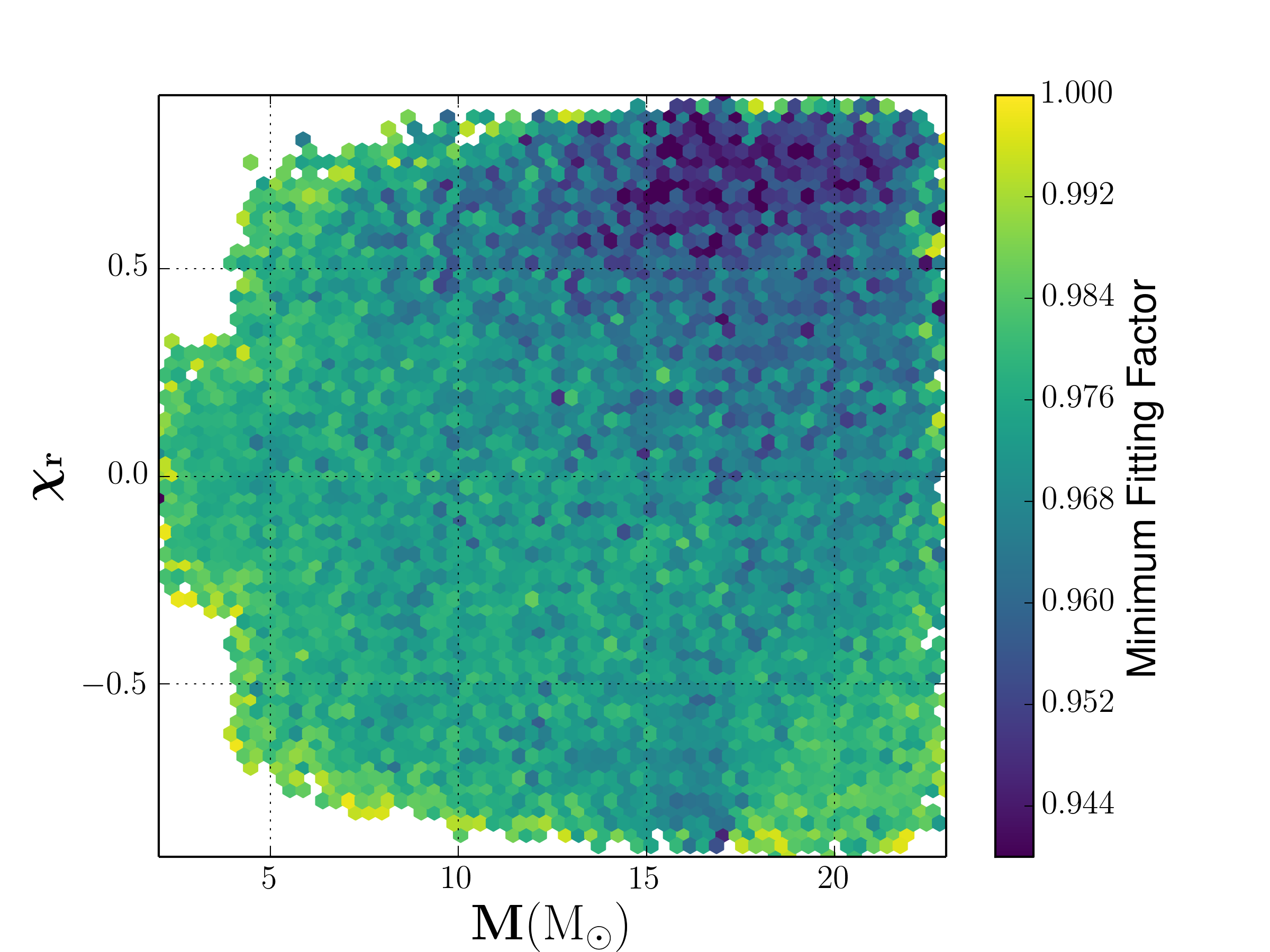}}}%
    					   
    \caption{The left figure shows the minimum fitting factor for the hybrid template bank for a set of injected aligned spin NSBH, BNS signals as a function of component masses. The figure on the right shows the minimum fitting factor as a function of the reduced spin parameter and the total mass. Both the signal and template waveforms are modeled using the $\tfrs$ approximant. We see a  region showing poor coverage corresponding to high total masses and high reduced spin values. This is also seen for the stochastic bank as well. One possible reason could be that the $\tfrs$ metric could be doing worse for systems with high masses and high reduced spin.}
    \label{fig:ff}
\end{figure}

In Fig.~\ref{fig:ff}\subref{fig_ff_a}--\subref{fig_ff_b},  we show the histogram of ``minimum'' fitting factors \cite{Harry-2014} (for geometric-stochastic template bank) over various combinations of intrinsic parameters of the compact binary system where both the signals and template waveforms are generated from $\tfrs$ waveform model. The bank performs well to match the injected signals throughout the parameter space for NSBH and BNS systems except for those regions where both the total mass and reduced spins are high.

%------------------

%The advantage of geometric-stochastic template placement is expected to be diluted for a smaller parameter space where the boundary effects may dominate. In order to test and quantify this hypothesis we generate another template bank using parameters in Set-II of Table \ref{tab:tmp}. We also compare this geometric-stochastic bank against the vanilla stochastic bank and show that they provide nearly equal coverage. 

In order to carry out a high-precision test comparing the efficiency of the hybrid bank with that of the vanilla-stochastic bank, we need a large number of injections to calculate the fitting factors upto high significant figures. To this end, 
we construct both the template banks using $\tfrs$ metric and parameters given in Set-II of Table \ref{tab:tmp}. The geometric random bank was constructed by initializing $\rlist$ with $1\times 10^7$ uniform random points in dimensionless chirp-time coordinates whereas the stochastic bank code was set to terminate when the rejection rate reached a value $0.9996$ averaged over the last $10$ acceptances.
The geometric random bank was found to contain $107,547$ templates whereas the stochastic bank contained ($25\%$ more) $134,563$ template points. 
%So as expected, we lose space-efficiency for smaller parameter space volume. 
In this case, the geometric-stochastic bank took $\gtrsim 11$ times less time than the vanilla-stochastic bank on a single unloaded processor. 
%In our numerical experiments, we have noticed that the new algorithm is at least an order of magnitude faster in placing the grid of templates for a wide range of parameters - this factor increases with the volume of the deemed parameter space and can go as high as $\times 50$ as reported in earlier section. In all these cases, the efficiency of the bank placed by the geometric-random method (as measured via fitting factors) is nearly equal to that placed using vanilla stochastic algorithm. 

\begin{figure}
\centering
  \includegraphics[width=0.6\textwidth]{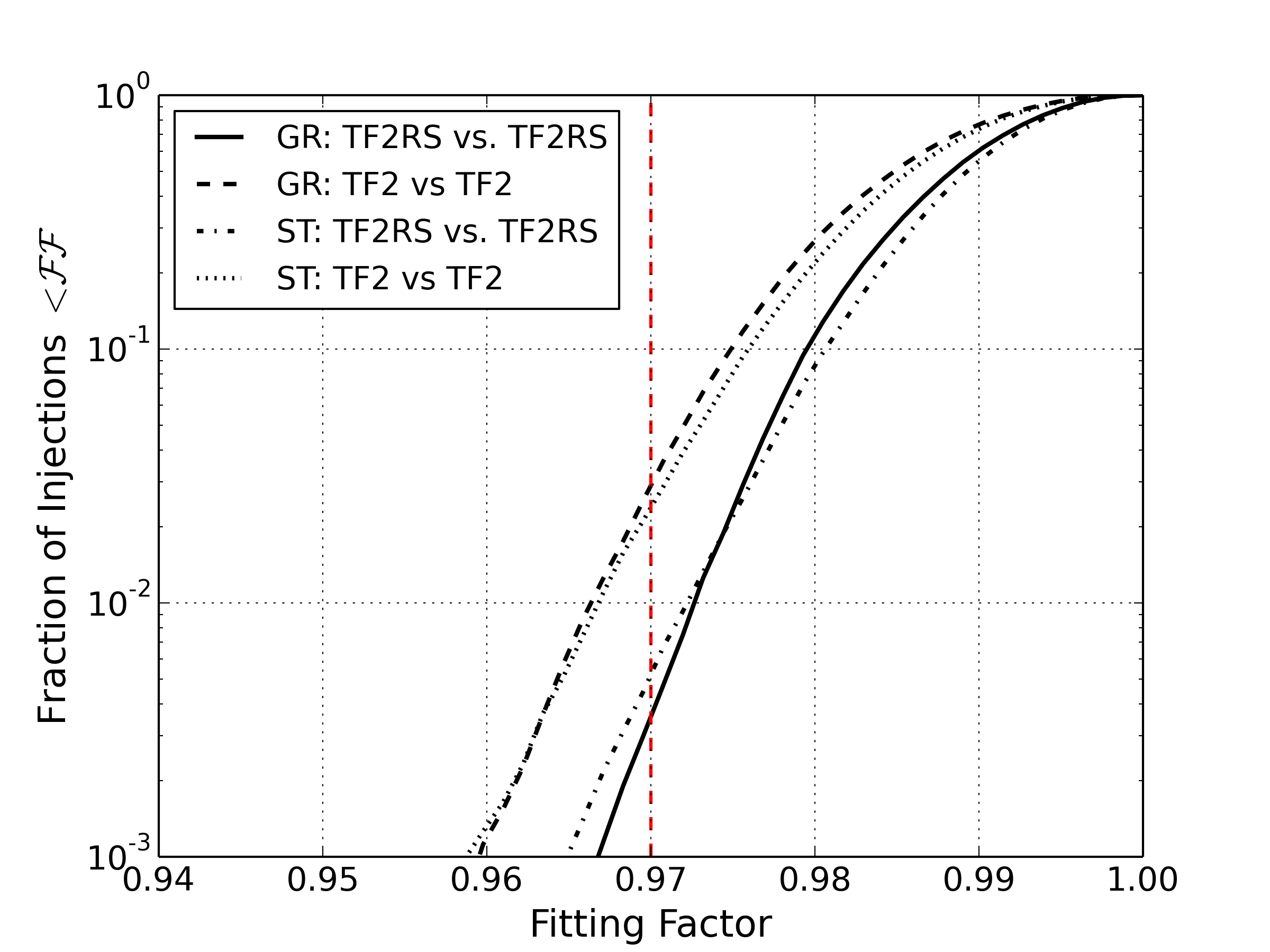}
  \caption{ 
  Fitting factors computed for various sets of aligned-spin signal families against geometric-random (GR) / vanilla stochastic template (ST) banks (see Set-II of Table~\ref{tab:tmp} for parameters). The performance of geometric random bank when both template and injected signals are generated from $\tfrs$ (TF2RS) approximant (black solid line), both generated from TaylorF2 (TF2) approximant (black dashed line) are shown. The performance of vanilla stochastic bank when both the templates injected signals are generated from $\tfrs$ approximant (black dashed dot line) and  when both are generated from TaylorF2 approximant (black dot line) are also shown. In this case, the vanilla stochastic bank has 25\% more templates than the geometric random bank and can be placed about $\gtrsim 11$ times faster in time.}
  \label{fig:tf2alrs}
\end{figure}

We quantify the performance of these two template banks by computing fitting factors for two different injection families of aligned spin waveforms: $\tfrs$ and TaylorF2. $50,000$ injections were made in both cases where the intrinsic parameters of the injected waveforms were chosen from Set-II of Table~\ref{tab:tmp} and other parameters chosen as earlier. For the case of geometric-random template bank created using local TO lattice, we found $0.354\%$  signals were recovered below  a fitting factor of 0.97 for $\tfrs$ injections, and $2.892\%$ signals are recovered below  this level for TaylorF2 injections. In the latter case, the injection model is different from the one used to construct the template bank - hence it is expected that the coverage for TaylorF2 will be less. The corresponding numbers for the vanilla-stochastic bank are found to be $0.514 \%$ ($\tfrs$ injections) and $2.408\%$ (TaylorF2 injections) respectively. From these numbers it is evident that the geometric-stochastic bank is equally efficient as the vanilla-stochastic bank. These results are summarized in Fig.~\ref{fig:tf2alrs}.

%-- A_3^* seed bank
As mentioned earlier, the $A_3^*$ lattice provides optimal coverage for conformally flat spaces in 3D. One may be tempted therefore to use such a lattice as a seed for stochastic template placement. We have investigated this approach and compared it with the geometric-random and vanilla-stochastic algorithms. At first, the deemed parameter space (corresponding to parameters in Set-II of Table \ref{tab:tmp}) was covered by a $A_3^*$ lattice using the metric at a putative point to determine the dimensions of the unit cell. Using a point in this volume for which the unit cell had the smallest dimension, we placed $31,732$ $A_3^*$ lattice points to entirely cover the deemed parameter space. Using these as seed points for stochastic placement, the final bank size was found to have $128,185$ templates. This is marginally ($\sim 5\%$) smaller than the vanilla-stochastic bank which has $134,563$ templates as reported above. The geometric-random bank outperforms this $A_3^*$ seeded stochastic bank by a good margin of more than $16\%$.

%-- top-down bank
As remarked earlier, the vanilla-stochastic algorithm can be cast in two different ways. The traditional {\textit{bottom-up}} approach has been implementation in the LSC Algorithm Library and has been used in this work for comparison with the geometric-random bank. In order to compare it head-to-head with the top-down approach we implemented it in software and ran it for the exact same parameters as given in Set II of Table~\ref{tab:tmp}. As expected, the top-down implementations gives nearly identical bank sizes ($0.1\%$ difference in size) but takes less than half the time as the bottom-up LAL implementation. The computational speed comes from the fact that in this implementation, one can reject many proposals that lie within the minimal match ellipsoid centred around a single accepted proposal. Efficient computational data structures like binary search trees are readily available for such queries. 

The summary of various templates banks referred to in the above discussion is available in Table \ref{tab:bankSummary}.

% \begin{table*}[htbp!]
% \begin{tabular}{|p{3cm}|p{3.55cm}|p{2cm}|p{2cm}|p{3.5cm}|} % {|l|l|r|r|l|} 
% \hline \hline
% Bank Parameters           &	Placement Algorithm & Bank Size     & Execution Time (min) & Comments \\ \hline 
% \multirow{2}{*}{Set-I of Table \ref{tab:tmp}}	    & Geometric-Random	  & $ \ 889,338 $		& $ \ \ 154$	& \multirow{2}{*}{\begin{tabular}[b]{@{}c@{}}  $35\%$ fewer templates \\$\times 51$ faster\end{tabular}} \\ \cline{2-4}
%     										  & Vanilla-Stochastic (\texttt{lalapps\_cbc\_sbank})	& $ \ 1,153,196$	& $ \ \ 7952$	 &	\\ 
%     
%     \hline
%     \hline
%     \multirow{4}{*}{Set-II of Table \ref{tab:tmp}}	& Geometric-Random	& $ \ 107,547$		  & $ \ \ 69$	 & \multirow{3}{*}{\begin{tabular}[b]{@{}c@{}}  $25\%$ fewer templates \\$\times 11$ faster \end{tabular}}\\ \cline{2-4}
%     										& Vanilla-Stochastic (\texttt{lalapps\_cbc\_sbank})	& $ \ 134,563$	& $ \ \ 762$ & \\ \cline{2-4}
%     										& $A_3^*$ seeded Vanilla-Stochastic	& $ \ 128,185$		& $\ \ --$ & \\ \cline{2-4}
%     										& Vanilla-Stochastic \hspace{2cm} (\texttt{top\_down}) & \ 134, 426 & $ \ \ 320$ & \multirow{1}{*}{\begin{tabular}[b]{@{}c@{}} $\times 2.3$ faster \end{tabular}} \\ \cline{2-4}
%     \hline
% \end{tabular}

\begin{table*}[htbp!]
\begin{tabular}{|p{3cm}|p{3.7cm}|p{1.5cm}|p{2cm}|p{2cm}|p{3.5cm}|} % {|l|l|r|r|l|} 
\hline \hline
Bank Parameters           &	Placement Algorithm & Size of $\rlist$ & Bank Size     & Execution Time (min) & Comments \\ \hline 
\multirow{8}{*}{Set-I of Table \ref{tab:tmp}}	    & &  $  1\times 10^7$	  & $ \ 694,422 $		& $ \ \ 375$	& \multirow{2}{*}{\begin{tabular}[b]{@{}c@{}}  $25\%$ more templates \\$\times (8-10)$ faster\end{tabular}} \\ \cline{3-5}
                          &  &  $ 2\times 10^7$	  & $ \ 749,705 $		& $ \ \ 482$	& \\ \cline{3-5}
                          & Hybrid construction &  $  3\times 10^7$	  & $ \ 777,113 $		& $ \ \ 616$	& \\ \cline{3-5}
                          & Geometric-Random &  $  4\times 10^7$	  & $ \ 798,269 $		& $ \ \ 885$	& \\ \cline{3-5}
                          &  &  $  5\times 10^7$	  & $ \ 812,570 $		& $ \ \ 990$	& \\ \cline{3-5}
                          &  &  $  6\times 10^7$	  & $ \ 824,541 $		& $ \ \ 1191$	& \\ \cline{3-5}
                          &  &  $  8\times 10^7$	  & $ \ 843,177 $		& $ \ \ 1712$	& \\ \cline{2-5}
    										  & Vanilla-Stochastic \footnotesize{(\texttt{lalapps\_cbc\_sbank})}	&  $--$ & $\ 939,787$	& $ \ \ 3666$	 &	\\ 
    
    \hline
    \hline
    \multirow{4}{*}{Set-II of Table \ref{tab:tmp}}	& Hybrid construction Geometric-Random	& $1\times 10^7$ &    $ \ 107,547$		& $ \ \ 69$	 & \multirow{3}{*}{\begin{tabular}[b]{@{}c@{}}  $25\%$ more templates \\$\times 11$ faster\end{tabular}}\\ \cline{2-5}
    										& Vanilla-Stochastic \footnotesize{(\texttt{lalapps\_cbc\_sbank})}	& $--$ & $ \ 134,563$	& $ \ \ 762$	 &	\\ \cline{2-5}
    										& $A_3^*$ seeded Stochastic	& $--$ & $ \ 128,185$		& $--$ & \\ \cline{2-5}
    										& Vanilla-Stochastic \hspace{2cm} \footnotesize(\texttt{top\_down}) & $--$ & \ 134, 426 & $ \ \ 320$ & \multirow{1}{*}{\begin{tabular}[b]{@{}c@{}} $\times 2.3$ faster \end{tabular}} \\ \cline{2-5}
    \hline
\end{tabular}

\caption{Summary of various template banks mentioned in this paper. The semi-analytic metric $g_{ij}$ for $\tfrs$ signal model was used in all cases. The Vanilla Stochastic algorithm can also work by directly calculating matches (instead of using $g_{ij}$) but in this head-to-head comparison, we have used the semi-analytic metric which speeds up the Vanilla Stochastic template placement significantly. The usage of the metric is compulsory for Geometric-Random placement. }
\label{tab:bankSummary}
\end{table*}

\section{Discussion and outlook}
\label{sec:Conclusions}

Templated matched filtering is the mainstay of gravitational wave detection pipelines. With unprecedented improvement in low frequency sensitivity of advanced detectors, and the availability of theoretical spinning waveform models, it has become imperative to conduct these searches over increasingly larger volumes in higher dimensional parameter spaces. For such cases, the stochastic algorithm is used for template placement as it is easily scalable to higher dimensions. But it is computationally expensive and by design not the most space efficient. 

This paper introduces a new template placement algorithm in 3D with an attempt to combine the space efficiency of $A_3^*$ lattice along with the robustness of stochastic placement algorithm. Such a template bank can be used in gravitational wave searches from binary neutron stars and neutron-star-blackhole compact binary systems where the waveform is described by two mass parameters and a mass-weighted spin magnitude parameter providing coverage for aligned-spin systems. 

The truncated octahedron which is a Dirichlet-Voronoi polytope of the $A_3^*$ lattice, inscribed within the minimal match ellipsoid is used as a unit cell for the geometric placement. Such lattices are known to provide optimal coverage for conformally flat 3D spaces. While the signal manifold is not globally flat, one can assume local flat patches and use such cells to cover them optimally. The interface to the stochastic placement algorithm is made by spraying random points over the parameters space - which are removed if found within the minimal match ellipsoid of any template. We discuss how this merger of methods is able to handle the issues arising out of varying curvature and irregular boundaries.
The nuances of its optimal implementation  are discussed in detail. We show in a direct comparison with stochastic algorithm that the new method generates significantly fewer templates and is computationally more efficient in Table\ref{tab:bankSummary}. We now make a couple of comments related to the new template placement algorithm presented in this paper.

One of the key issues with the geometric template placement algorithms (e.g. geometric hexagonal placement \cite{Cokelaer-2007}) is related to the amount of \textit{fine tuning} needed in the method to account for curvature effects. For example, in 2D the hexagonal cells used to place templates need to be whittled or pared to a slightly smaller size so that no holes are left due to the changes in relative orientation of neighboring hexagonal cells. This reduces the overall efficiency of placement leading to increase in template bank sizes. But more importantly, one loses the ability to make generic codes that are usable for different waveform models. In this regard, the geometric-random algorithm proposed here is robust against such fine-tuning by design. The template placement proceeds by first spraying a large number of random points over the parameters space which are later removed if found within the minimal match ellipsoid of the templates in the bank. Suppose a small portion of the deemed space is left uncovered due to curvature effects, it would lead to some residual points that are not removed from $\rlist$. As evident from Algorithm \ref{algo:gr}, these points are revisited in subsequent iterations where a random point (out of the residuals) is added to the template bank leading to complete coverage.  

We would like to point out that the template banks shown in this paper make use of the $\tfrs$ waveform family which model only the inspiral part of the evolution of the compact objects. They are not ideal for BBH searches directly, where a significant part of the SNR is expected to be contributed from the merger and ringdown phases. However the method developed in this paper is quite general and can be used for placing templates for IMR waveform families as well. If analytical metrics for IMR approximants (IMRPhenomD\cite{Khan-2016, Husa-2016}, SEOBNRv2\cite{Taracchini-2014, Purrer-2014}, etc.) were available, the proposed method would be an elegant and efficient solution for covering the entire combined BNS+NSBH+BBH space. This will thereby mesh well with modern LIGO searches which are carried out over a large range of parameters with combined BNS+NSBH+BBH banks using a mix of inspiral- only and IMR templates.

The assumption that the metric on the signal manifold is slowly varying and is locally flat is crucial for space efficiency. One can imagine a hypothetical case where this assumption does not hold true (e.g. metric coefficients are random at every point) in which case, the geometric-random algorithm will effectively fall back to the top-down version of the vanilla-stochastic template placement by design. In other words, the number of templates in the bank from the new method will not exceed the stochastic template bank in the limiting case. We have also shown in a direct comparison that the top-down stochastic bank implementation is computationally more efficient and should be used where the metric is available. Incorporating an intelligent way of spraying the random proposals (instead of drawing them from a uniform distribution) over the parameter space may lead to further optimization of this method.

On the other hand if the metric was perfectly flat (and given in coordinates where it was constant), 
the hybrid construction would fall back on a perfect $A_3^*$ lattice.

Finally, the geometric-random placement method presented here for 3D is generically scalable to signal manifolds in higher dimensions by using the appropriate $A_n^* \; (n > 3)$ lattices. 

%--------------------------------------------

\appendix

\section{The \texorpdfstring{$\tfrs$}{} signal model for gravitational waves from inspiraling compact binary coalescence}
\label{appendix:sigmodel}

The TaylorF2 “reduced spin” waveform model in frequency domain is given by\cite{Ajith-2011b}

\begin{equation}\label{eq:3.3}
 \widetilde{h}(f) = \mathcal{A}f^{-7/6}\exp\Big\{-i\big[\Psi(f) - \frac{\pi}{4}\big]\Big\}
\end{equation}

where, the amplitude $\mathcal{A}$ depends on the component masses, distance to the source, sky position, orientation of the binary's plane, and $\Psi(f)$ is the instantaneous phase which can be explicitly written as:

\begin{widetext}
\begin{equation}
\begin{split}
 \Psi(f) = 2\pi ft_{0} + \phi_{0} + \frac{3}{128\eta v_{f}^{5}}\Bigg\{ 1 + v_{f}^{2}\Big[\frac{55\eta}{9} + \frac{3715}{756}\Big]
 +v_{f}^{3}\big[4\beta - 16\pi\big]  + 
 v_{f}^{4}\Big[\frac{3085\eta^{2}}{72} +  \frac{27145\eta}{504} \\
 + \frac{15293365}{508032}
 -10\sigma_{0}\Big]  + v_{f}^{5}\Big[\frac{38645\pi}{756} -\frac{65\pi}{\eta} - \gamma_{0}\Big]\big(3ln(v_{f}) + 1\big) + 
 v_{f}^{6}\Big[-\frac{6848\gamma_{E}}{21} - \frac{127825\eta^{3}}{1296} + \\
 \frac{76055\eta^{2}}{1728} +
 \big(\frac{2255\pi^{2}}{12}
 - \frac{15737765635}{3048192}\big)\eta -\frac{640\pi^{2}}{3} + \frac{11583231236531}{4694215680}-
 \frac{6848ln(4v_{f})}{21}\Big] + \\
 v_{f}^{7}\Big[-\frac{74045\pi\eta^{2}}{756} + \frac{37815\pi\eta}{1512} + 
 \frac{77096675\pi}{254016}\Big] \Bigg\}
\end{split}
\end{equation}
\end{widetext}

\noindent where $t_{0}$ is the time of arrival of the signal at the detector marking the epoch at which the instantaneous frequency takes a fiducial value, $\phi_{0}$ is the corresponding phase,  $v_{f} \equiv (\pi m f)^{1/3}$ is the instantaneous velocity, $m = m_{1} + m_{2}$ is the total mass and $\eta = m_{1}m_{2}/(m_{1} + m_{2})^{2}$ is the symmetric mass ratio of the binary and $\gamma_{E}$ is the Euler gamma constant.

The spin effects are encoded through  $\beta$, $\sigma_{0}$ and $\gamma_{0}$ which appear at 1.5PN, 2PN and 2.5PN phase terms respectively and are given by:
\begin{equation}
 \begin{split}
  \beta &= \frac{113\chi_{r}}{12} \\
  \sigma_{0} &= \Big(-\frac{12769(4\eta -81)}{16(76\eta -113)^{2}}\Big)\chi_r^{2}\\
  \gamma_{0} &= \Big(\frac{565(1713\eta^{2} + 135856\eta - 146597)}{2268(76\eta -13)}\Big) \chi_r
 \end{split}
\end{equation}
\noindent where the reduced spin parameter $\chi_{r}$ is defined as the weighted sum of individual spins $\chi_{1}$ and $\chi_{2}$ of the component masses as:
\begin{equation}
 \chi_{r} = \frac{1}{2}\Big(1 - \frac{76}{113\eta}\Big)(\chi_{1} + \chi_{2}) + \frac{1}{2}\frac{m_{1} - m_{2}}{m_{1} + m_{2}}(\chi_{1} - \chi_{2})
\end{equation}

\noindent The individual spins of the components $\chi_{1,2}$ are the projections of their spin vectors $\boldsymbol{S}_{1,2}$ along the Newtonian orbital angular momentum vector \textbf{$\boldsymbol{{L}}_{N}$} and defined as: 
\begin{equation}
 \chi_{1,2} = \frac{\boldsymbol{S}_{1,2}\cdot \boldsymbol{\hat{L}}_{N}}{2m_{1,2}^{2}}.
\end{equation}

\section{Space filling truncated octahedron}
\label{appendix:TO}

A polyhedron is a three dimensional solid that has a finite number of polygon faces. One can fill a 3D space completely without any overlap or gap through the tessellations of space-filling polyhedra. Examples of such space-filling polyhedron include cube, hexagonal prism etc. Solution to such space-filling problems find many practical applications like optimal placement of a network of communication towers \citep{spf}. The template placement problem addressed in this paper can also be mapped to an optimal space-filling problem in curved space. The geometric properties of optimal space-filling polyhedra can be understood from the volumetric quotient $(Q_{v})$ defined as
 \begin{equation}
  Q_{v} = \frac{V}{\frac{4}{3}\pi R^3},
  \label{eq:Qv}
 \end{equation}
where $V$ is the volume of polyhedron, $R$ is maximum distance from its center to any vertex and $S$ is the surface area of the polyhedron. Note that $(Q_v)$ is the inverse of the \emph{thickness}, a common measure for the efficiency of a given covering. The polyhedra with the highest $Q_v$ is expected to provide the optimum coverage.
%The optimal space-filling polyhedron is expected to have a high value of $Q_i$. The isoperimetric quotient of a truncated-octahedron is  $0.753$ while that for a WP is marginally higher at $0.764$.  

\begin{figure}%
    \centering
 {\includegraphics[width=0.4\textwidth]{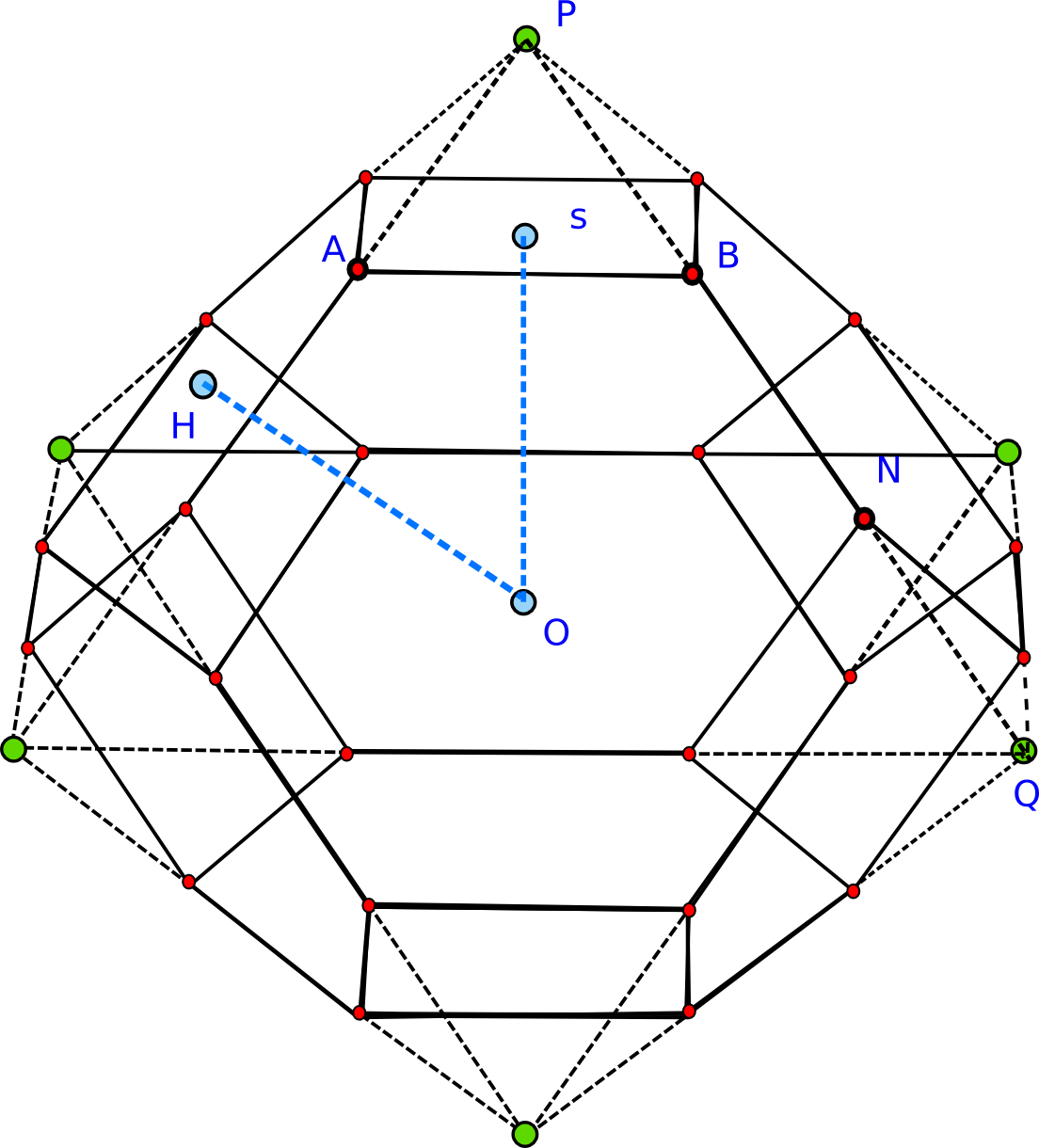}}   					   
    \caption{Truncated Octahedron created by truncating six pyramids from the six vertices of two square base pyramids.}
    \label{fig:to}%
\end{figure}

The TO is a 14-faced  Archimedean solid, with 8 hexagonal faces, 6 square faces and has 24 vertices. It is generated by joining two regular pyramids upside down and cutting a pyramid from all six vertices in such a way that the length of all the sides generated are equal. Thus a truncated octahedron of side $a$ can be created by removing six pyramids of side $a$ from an octahedron of side $3a$. Fig. \ref{fig:to} shows the TO obtained from two pyramids. The geometric properties of a TO and pyramid are given in Table \ref{tab:togeom}.

\begin{table}
 \centering
  \begin{tabular}{ p{7.0cm} | p{2.7cm}  }
 \hline \hline
 Name  &  Value\\
 \hline \hline
 Side length  of Pyramid $(PQ)$           & $3a$\\
 Side length of TO $(BN)$  				        & $a$\\
 Height of Pyramid $(OP)$ 				        & $\frac{3}{\sqrt{2}}a$\\
 Height of TO along square face $(OS)$ 	  & $\sqrt{\frac{3}{2}}a$\\\ 
 Height of TO along hexagonal face $(OH)$ & $\sqrt{2}a$\\
 Height of TO along vertices $(ON)$ 		  & $\sqrt{\frac{5}{2}}a$\\
 Volume of TO $(V)$  						          & $8\sqrt{2}a^3$ \\
 Surface Area of TO $(S)$  					      & $(6+12\sqrt{3})a^2 $\\
 Volumetric Quotient of TO $(Q_{v})$ 		  & $\approx 0.683292042 $\\
 %Isoperimetric Quotient of TO $(Q_{i})$ 	& $\approx 0.753366625$ \\
 \hline
\end{tabular}
\caption{Geometrical properties of truncated octahedron and pyramid where the truncated octahedron constructed using two truncated pyramid.}
\label{tab:togeom}
\end{table}

Suppose a three dimensional volume is covered by the tessellations of TO cells. Each face of such a cell can be shared by  another neighboring TO cell. There are two kind of neighboring cells: ones that share the square faces which we shall call S-neighbors and the others that share the hexagonal faces which we shall refer to as H-neighbors. Each TO has a maximum of six S-neighbors and eight H-neighbors. The distance between S-neighbors is twice the height of TO along square face and for H-neighbors, twice of height of TO along the hexagonal face. When a TO is inscribed in a sphere of radius $R$ such that the $z$-axis goes through the centre of one of the square faces, then, the sides of the squares and hexagons $a$ is given by $a = \sqrt{\frac{2}{5}}R$, the distance from the centre to each of the S-neighbors is equal to $\sqrt{\frac{12}{5}} R$, and the distance to the H-neighbors is $\frac{4}{\sqrt{5}}R$. The coordinates of all the 14 neighbours are listed in Table~\ref{tab:tongb}.

\begin{table}
 \begin{tabular}{ p{3cm} | p{5cm}  }
 \hline
 \hline
 Neighborhood Type  &  Position\\
 \hline
 \hline
 S-Neighborhood & $(\frac{2R}{\sqrt{5}}, \frac{2R}{\sqrt{5}}, \frac{2R}{\sqrt{5}}), 
 (-\frac{2R}{\sqrt{5}}, -\frac{2R}{\sqrt{5}}, - \frac{2R}{\sqrt{5}}),$ \\
 
  square faces& $ (-\frac{2R}{\sqrt{5}}, \frac{2R}{\sqrt{5}}, -\frac{2R}{\sqrt{5}}),
   (\frac{2R}{\sqrt{5}}, -\frac{2R}{\sqrt{5}}, -\frac{2R}{\sqrt{5}})$, \\
   
   &$(-\frac{2R}{\sqrt{5}}, \frac{2R}{\sqrt{5}}, \frac{2R}{\sqrt{5}}), 
   (\frac{2R}{\sqrt{5}}, \frac{2R}{\sqrt{5}}, -\frac{2R}{\sqrt{5}}),$\\
   
   & $(-\frac{2R}{\sqrt{5}}, -\frac{2R}{\sqrt{5}}, \frac{2R}{\sqrt{5}}), 
   (\frac{2R}{\sqrt{5}}, -\frac{2R}{\sqrt{5}}, \frac{2R}{\sqrt{5}})$\\
 & \\
 H-Neighborhood & $(0, \frac{4R}{\sqrt{5}}, 0), (0, -\frac{4R}{\sqrt{5}}, 0), (\frac{4R}{\sqrt{5}}, 0, 0),$\\
    hexagonal faces & $(-\frac{4R}{\sqrt{5}}, 0, 0), (0, 0, \frac{4R}{\sqrt{5}}), (0, 0, -\frac{4R}{\sqrt{5}})$\\
    
 \hline
 
 \hline
\end{tabular}
\caption{These coordinates are the neighborhood positions of a TO where a TO is inscribed in a sphere with radius $R$, where the center of a TO is placed at origin and the $z$-axis goes through the centre of one of the square faces.}
\label{tab:tongb}
\end{table}

\begin{acknowledgments}
A.S. would like to thank LS for motivation. SR thanks IIT Gandhinagar for research fellowship and fellow graduate students (Amit Reza, Chakresh Singh, Md. Yousuf) for useful discussions and help with the manuscript. We thank Samarth Vaijanapurkar (B.Tech. 2016 student of IIT Gandhinagar) for useful discussions. The authors thank Duncan Brown for valuable feedback. Last but not least, the authors thank the anonymous referees whose careful reading of the manuscript, detailed queries and many suggestions have helped to improve the content and presentation of this work.
\end{acknowledgments}

%\nocite{*}
%\nocite{apsrev41Control}
\bibliography{reference}{}

%merlin.mbs apsrev4-1.bst 2010-07-25 4.21a (PWD, AO, DPC) hacked
%Control: key (0)
%Control: author (72) initials jnrlst
%Control: editor formatted (1) identically to author
%Control: production of article title (-1) disabled
%Control: page (0) single
%Control: year (1) truncated
%Control: production of eprint (0) enabled
\begin{thebibliography}{52}%
\makeatletter
\providecommand \@ifxundefined [1]{%
 \@ifx{#1\undefined}
}%
\providecommand \@ifnum [1]{%
 \ifnum #1\expandafter \@firstoftwo
 \else \expandafter \@secondoftwo
 \fi
}%
\providecommand \@ifx [1]{%
 \ifx #1\expandafter \@firstoftwo
 \else \expandafter \@secondoftwo
 \fi
}%
\providecommand \natexlab [1]{#1}%
\providecommand \enquote  [1]{``#1''}%
\providecommand \bibnamefont  [1]{#1}%
\providecommand \bibfnamefont [1]{#1}%
\providecommand \citenamefont [1]{#1}%
\providecommand \href@noop [0]{\@secondoftwo}%
\providecommand \href [0]{\begingroup \@sanitize@url \@href}%
\providecommand \@href[1]{\@@startlink{#1}\@@href}%
\providecommand \@@href[1]{\endgroup#1\@@endlink}%
\providecommand \@sanitize@url [0]{\catcode `\\12\catcode `\$12\catcode
  `\&12\catcode `\#12\catcode `\^12\catcode `\_12\catcode `\%12\relax}%
\providecommand \@@startlink[1]{}%
\providecommand \@@endlink[0]{}%
\providecommand \url  [0]{\begingroup\@sanitize@url \@url }%
\providecommand \@url [1]{\endgroup\@href {#1}{\urlprefix }}%
\providecommand \urlprefix  [0]{URL }%
\providecommand \Eprint [0]{\href }%
\providecommand \doibase [0]{http://dx.doi.org/}%
\providecommand \selectlanguage [0]{\@gobble}%
\providecommand \bibinfo  [0]{\@secondoftwo}%
\providecommand \bibfield  [0]{\@secondoftwo}%
\providecommand \translation [1]{[#1]}%
\providecommand \BibitemOpen [0]{}%
\providecommand \bibitemStop [0]{}%
\providecommand \bibitemNoStop [0]{.\EOS\space}%
\providecommand \EOS [0]{\spacefactor3000\relax}%
\providecommand \BibitemShut  [1]{\csname bibitem#1\endcsname}%
\let\auto@bib@innerbib\@empty
%</preamble>
\bibitem [{\citenamefont {Abbott}\ \emph
  {et~al.}(2016{\natexlab{a}})\citenamefont {Abbott} \emph
  {et~al.}}]{gw150914}%
  \BibitemOpen
  \bibfield  {author} {\bibinfo {author} {\bibfnamefont {B.~P.}\ \bibnamefont
  {Abbott}} \emph {et~al.} (\bibinfo {collaboration} {LIGO Scientific
  Collaboration and Virgo Collaboration}),\ }\href {\doibase
  10.1103/PhysRevLett.116.061102} {\bibfield  {journal} {\bibinfo  {journal}
  {Phys. Rev. Lett.}\ }\textbf {\bibinfo {volume} {116}},\ \bibinfo {pages}
  {061102} (\bibinfo {year} {2016}{\natexlab{a}})}\BibitemShut {NoStop}%
\bibitem [{\citenamefont {Harry}\ and\ \citenamefont {the LIGO
  Scientific~Collaboration}(2010)}]{advancedLIGO}%
  \BibitemOpen
  \bibfield  {author} {\bibinfo {author} {\bibfnamefont {G.~M.}\ \bibnamefont
  {Harry}}\ and\ \bibinfo {author} {\bibnamefont {the LIGO
  Scientific~Collaboration}},\ }\href
  {http://stacks.iop.org/0264-9381/27/i=8/a=084006} {\bibfield  {journal}
  {\bibinfo  {journal} {Classical and Quantum Gravity}\ }\textbf {\bibinfo
  {volume} {27}},\ \bibinfo {pages} {084006} (\bibinfo {year}
  {2010})}\BibitemShut {NoStop}%
\bibitem [{\citenamefont {Abbott}\ \emph
  {et~al.}(2016{\natexlab{b}})\citenamefont {Abbott} \emph
  {et~al.}}]{gw151226}%
  \BibitemOpen
  \bibfield  {author} {\bibinfo {author} {\bibfnamefont {B.~P.}\ \bibnamefont
  {Abbott}} \emph {et~al.} (\bibinfo {collaboration} {LIGO Scientific
  Collaboration and Virgo Collaboration}),\ }\href {\doibase
  10.1103/PhysRevLett.116.241103} {\bibfield  {journal} {\bibinfo  {journal}
  {Phys. Rev. Lett.}\ }\textbf {\bibinfo {volume} {116}},\ \bibinfo {pages}
  {241103} (\bibinfo {year} {2016}{\natexlab{b}})}\BibitemShut {NoStop}%
\bibitem [{\citenamefont {Acernese}\ \emph {et~al.}(2015)\citenamefont
  {Acernese} \emph {et~al.}}]{advancedVIRGO-2015}%
  \BibitemOpen
  \bibfield  {author} {\bibinfo {author} {\bibfnamefont {F.}~\bibnamefont
  {Acernese}} \emph {et~al.},\ }\href
  {http://stacks.iop.org/1742-6596/610/i=1/a=012014} {\bibfield  {journal}
  {\bibinfo  {journal} {Journal of Physics: Conference Series}\ }\textbf
  {\bibinfo {volume} {610}},\ \bibinfo {pages} {012014} (\bibinfo {year}
  {2015})}\BibitemShut {NoStop}%
\bibitem [{\citenamefont {Somiya}(2012)}]{kagra-2012}%
  \BibitemOpen
  \bibfield  {author} {\bibinfo {author} {\bibfnamefont {K.}~\bibnamefont
  {Somiya}},\ }\href {http://stacks.iop.org/0264-9381/29/i=12/a=124007}
  {\bibfield  {journal} {\bibinfo  {journal} {Classical and Quantum Gravity}\
  }\textbf {\bibinfo {volume} {29}},\ \bibinfo {pages} {124007} (\bibinfo
  {year} {2012})}\BibitemShut {NoStop}%
\bibitem [{\citenamefont {Aso}\ \emph {et~al.}(2013)\citenamefont {Aso},
  \citenamefont {Michimura}, \citenamefont {Somiya}, \citenamefont {Ando},
  \citenamefont {Miyakawa}, \citenamefont {Sekiguchi}, \citenamefont
  {Tatsumi},\ and\ \citenamefont {Yamamoto}}]{kagra-2013}%
  \BibitemOpen
  \bibfield  {author} {\bibinfo {author} {\bibfnamefont {Y.}~\bibnamefont
  {Aso}}, \bibinfo {author} {\bibfnamefont {Y.}~\bibnamefont {Michimura}},
  \bibinfo {author} {\bibfnamefont {K.}~\bibnamefont {Somiya}}, \bibinfo
  {author} {\bibfnamefont {M.}~\bibnamefont {Ando}}, \bibinfo {author}
  {\bibfnamefont {O.}~\bibnamefont {Miyakawa}}, \bibinfo {author}
  {\bibfnamefont {T.}~\bibnamefont {Sekiguchi}}, \bibinfo {author}
  {\bibfnamefont {D.}~\bibnamefont {Tatsumi}}, \ and\ \bibinfo {author}
  {\bibfnamefont {H.}~\bibnamefont {Yamamoto}} (\bibinfo {collaboration} {The
  KAGRA Collaboration}),\ }\href {\doibase 10.1103/PhysRevD.88.043007}
  {\bibfield  {journal} {\bibinfo  {journal} {Phys. Rev. D}\ }\textbf {\bibinfo
  {volume} {88}},\ \bibinfo {pages} {043007} (\bibinfo {year}
  {2013})}\BibitemShut {NoStop}%
\bibitem [{\citenamefont {Iyer}\ \emph {et~al.}(2011)\citenamefont {Iyer},
  \citenamefont {Souradeep}, \citenamefont {Unnikrishnan}, \citenamefont
  {Dhurandhar}, \citenamefont {Raja},\ and\ \citenamefont
  {Sengupta}}]{indigo-proposal-2011}%
  \BibitemOpen
  \bibfield  {author} {\bibinfo {author} {\bibfnamefont {B.}~\bibnamefont
  {Iyer}}, \bibinfo {author} {\bibfnamefont {T.}~\bibnamefont {Souradeep}},
  \bibinfo {author} {\bibfnamefont {C.}~\bibnamefont {Unnikrishnan}}, \bibinfo
  {author} {\bibfnamefont {S.}~\bibnamefont {Dhurandhar}}, \bibinfo {author}
  {\bibfnamefont {S.}~\bibnamefont {Raja}}, \ and\ \bibinfo {author}
  {\bibfnamefont {A.}~\bibnamefont {Sengupta}},\ }\href {\doibase
  https://dcc.ligo.org/cgi-bin/DocDB/ShowDocument?docid=75988} {\bibfield
  {journal} {\bibinfo  {journal} {LIGO-India Technical Report}\ } (\bibinfo
  {year} {2011}),\
  https://dcc.ligo.org/cgi-bin/DocDB/ShowDocument?docid=75988}\BibitemShut
  {NoStop}%
\bibitem [{\citenamefont {Unnikrishnan}(2013)}]{indigo-unnikrishnan}%
  \BibitemOpen
  \bibfield  {author} {\bibinfo {author} {\bibfnamefont {C.~S.}\ \bibnamefont
  {Unnikrishnan}},\ }\href {\doibase 10.1142/S0218271813410101} {\bibfield
  {journal} {\bibinfo  {journal} {International Journal of Modern Physics D}\
  }\textbf {\bibinfo {volume} {22}},\ \bibinfo {pages} {1341010} (\bibinfo
  {year} {2013})}\BibitemShut {NoStop}%
\bibitem [{\citenamefont {Manca}\ and\ \citenamefont
  {Vallisneri}(2010)}]{coverart}%
  \BibitemOpen
  \bibfield  {author} {\bibinfo {author} {\bibfnamefont {G.~M.}\ \bibnamefont
  {Manca}}\ and\ \bibinfo {author} {\bibfnamefont {M.}~\bibnamefont
  {Vallisneri}},\ }\href {\doibase 10.1103/PhysRevD.81.024004} {\bibfield
  {journal} {\bibinfo  {journal} {Phys. Rev. D}\ }\textbf {\bibinfo {volume}
  {81}},\ \bibinfo {pages} {024004} (\bibinfo {year} {2010})}\BibitemShut
  {NoStop}%
\bibitem [{\citenamefont {Messenger}\ \emph {et~al.}(2009)\citenamefont
  {Messenger}, \citenamefont {Prix},\ and\ \citenamefont {Papa}}]{Prix-2009}%
  \BibitemOpen
  \bibfield  {author} {\bibinfo {author} {\bibfnamefont {C.}~\bibnamefont
  {Messenger}}, \bibinfo {author} {\bibfnamefont {R.}~\bibnamefont {Prix}}, \
  and\ \bibinfo {author} {\bibfnamefont {M.~A.}\ \bibnamefont {Papa}},\ }\href
  {\doibase 10.1103/PhysRevD.79.104017} {\bibfield  {journal} {\bibinfo
  {journal} {Phys. Rev. D}\ }\textbf {\bibinfo {volume} {79}},\ \bibinfo
  {pages} {104017} (\bibinfo {year} {2009})}\BibitemShut {NoStop}%
\bibitem [{\citenamefont {Cokelaer}(2007)}]{Cokelaer-2007}%
  \BibitemOpen
  \bibfield  {author} {\bibinfo {author} {\bibfnamefont {T.}~\bibnamefont
  {Cokelaer}},\ }\href {\doibase 10.1103/PhysRevD.76.102004} {\bibfield
  {journal} {\bibinfo  {journal} {Phys. Rev. D}\ }\textbf {\bibinfo {volume}
  {76}},\ \bibinfo {pages} {102004} (\bibinfo {year} {2007})}\BibitemShut
  {NoStop}%
\bibitem [{\citenamefont {Babak}\ \emph {et~al.}(2006)\citenamefont {Babak},
  \citenamefont {Balasubramanian}, \citenamefont {Churches}, \citenamefont
  {Cokelaer},\ and\ \citenamefont {Sathyaprakash}}]{Babak-2006}%
  \BibitemOpen
  \bibfield  {author} {\bibinfo {author} {\bibfnamefont {S.}~\bibnamefont
  {Babak}}, \bibinfo {author} {\bibfnamefont {R.}~\bibnamefont
  {Balasubramanian}}, \bibinfo {author} {\bibfnamefont {D.}~\bibnamefont
  {Churches}}, \bibinfo {author} {\bibfnamefont {T.}~\bibnamefont {Cokelaer}},
  \ and\ \bibinfo {author} {\bibfnamefont {B.~S.}\ \bibnamefont
  {Sathyaprakash}},\ }\href {http://stacks.iop.org/0264-9381/23/i=18/a=002}
  {\bibfield  {journal} {\bibinfo  {journal} {Classical and Quantum Gravity}\
  }\textbf {\bibinfo {volume} {23}},\ \bibinfo {pages} {5477} (\bibinfo {year}
  {2006})}\BibitemShut {NoStop}%
\bibitem [{\citenamefont {Harry}\ \emph {et~al.}(2009)\citenamefont {Harry},
  \citenamefont {Allen},\ and\ \citenamefont {Sathyaprakash}}]{Harry-2009}%
  \BibitemOpen
  \bibfield  {author} {\bibinfo {author} {\bibfnamefont {I.~W.}\ \bibnamefont
  {Harry}}, \bibinfo {author} {\bibfnamefont {B.}~\bibnamefont {Allen}}, \ and\
  \bibinfo {author} {\bibfnamefont {B.~S.}\ \bibnamefont {Sathyaprakash}},\
  }\href {\doibase 10.1103/PhysRevD.80.104014} {\bibfield  {journal} {\bibinfo
  {journal} {Phys. Rev. D}\ }\textbf {\bibinfo {volume} {80}},\ \bibinfo
  {pages} {104014} (\bibinfo {year} {2009})}\BibitemShut {NoStop}%
\bibitem [{\citenamefont {Brown}\ \emph {et~al.}(2012)\citenamefont {Brown},
  \citenamefont {Harry}, \citenamefont {Lundgren},\ and\ \citenamefont
  {Nitz}}]{Brown-2012}%
  \BibitemOpen
  \bibfield  {author} {\bibinfo {author} {\bibfnamefont {D.~A.}\ \bibnamefont
  {Brown}}, \bibinfo {author} {\bibfnamefont {I.}~\bibnamefont {Harry}},
  \bibinfo {author} {\bibfnamefont {A.}~\bibnamefont {Lundgren}}, \ and\
  \bibinfo {author} {\bibfnamefont {A.~H.}\ \bibnamefont {Nitz}},\ }\href
  {\doibase 10.1103/PhysRevD.86.084017} {\bibfield  {journal} {\bibinfo
  {journal} {Phys. Rev. D}\ }\textbf {\bibinfo {volume} {86}},\ \bibinfo
  {pages} {084017} (\bibinfo {year} {2012})}\BibitemShut {NoStop}%
\bibitem [{\citenamefont {Buonanno}\ \emph {et~al.}(2009)\citenamefont
  {Buonanno}, \citenamefont {Iyer}, \citenamefont {Ochsner}, \citenamefont
  {Pan},\ and\ \citenamefont {Sathyaprakash}}]{Buonanno-2009}%
  \BibitemOpen
  \bibfield  {author} {\bibinfo {author} {\bibfnamefont {A.}~\bibnamefont
  {Buonanno}}, \bibinfo {author} {\bibfnamefont {B.~R.}\ \bibnamefont {Iyer}},
  \bibinfo {author} {\bibfnamefont {E.}~\bibnamefont {Ochsner}}, \bibinfo
  {author} {\bibfnamefont {Y.}~\bibnamefont {Pan}}, \ and\ \bibinfo {author}
  {\bibfnamefont {B.~S.}\ \bibnamefont {Sathyaprakash}},\ }\href {\doibase
  10.1103/PhysRevD.80.084043} {\bibfield  {journal} {\bibinfo  {journal} {Phys.
  Rev. D}\ }\textbf {\bibinfo {volume} {80}},\ \bibinfo {pages} {084043}
  (\bibinfo {year} {2009})}\BibitemShut {NoStop}%
\bibitem [{\citenamefont {Poisson}\ and\ \citenamefont
  {Will}(1995)}]{Poisson-1995}%
  \BibitemOpen
  \bibfield  {author} {\bibinfo {author} {\bibfnamefont {E.}~\bibnamefont
  {Poisson}}\ and\ \bibinfo {author} {\bibfnamefont {C.~M.}\ \bibnamefont
  {Will}},\ }\href {\doibase 10.1103/PhysRevD.52.848} {\bibfield  {journal}
  {\bibinfo  {journal} {Phys. Rev. D}\ }\textbf {\bibinfo {volume} {52}},\
  \bibinfo {pages} {848} (\bibinfo {year} {1995})}\BibitemShut {NoStop}%
\bibitem [{\citenamefont {Ajith}\ \emph {et~al.}(2014)\citenamefont {Ajith},
  \citenamefont {Fotopoulos}, \citenamefont {Privitera}, \citenamefont
  {Neunzert}, \citenamefont {Mazumder},\ and\ \citenamefont
  {Weinstein}}]{Ajith-2014}%
  \BibitemOpen
  \bibfield  {author} {\bibinfo {author} {\bibfnamefont {P.}~\bibnamefont
  {Ajith}}, \bibinfo {author} {\bibfnamefont {N.}~\bibnamefont {Fotopoulos}},
  \bibinfo {author} {\bibfnamefont {S.}~\bibnamefont {Privitera}}, \bibinfo
  {author} {\bibfnamefont {A.}~\bibnamefont {Neunzert}}, \bibinfo {author}
  {\bibfnamefont {N.}~\bibnamefont {Mazumder}}, \ and\ \bibinfo {author}
  {\bibfnamefont {A.~J.}\ \bibnamefont {Weinstein}},\ }\href {\doibase
  10.1103/PhysRevD.89.084041} {\bibfield  {journal} {\bibinfo  {journal} {Phys.
  Rev. D}\ }\textbf {\bibinfo {volume} {89}},\ \bibinfo {pages} {084041}
  (\bibinfo {year} {2014})}\BibitemShut {NoStop}%
\bibitem [{\citenamefont {Capano}\ \emph {et~al.}(2016)\citenamefont {Capano},
  \citenamefont {Harry}, \citenamefont {Privitera},\ and\ \citenamefont
  {Buonanno}}]{Capano-2016}%
  \BibitemOpen
  \bibfield  {author} {\bibinfo {author} {\bibfnamefont {C.}~\bibnamefont
  {Capano}}, \bibinfo {author} {\bibfnamefont {I.}~\bibnamefont {Harry}},
  \bibinfo {author} {\bibfnamefont {S.}~\bibnamefont {Privitera}}, \ and\
  \bibinfo {author} {\bibfnamefont {A.}~\bibnamefont {Buonanno}},\ }\href
  {\doibase 10.1103/PhysRevD.93.124007} {\bibfield  {journal} {\bibinfo
  {journal} {Phys. Rev. D}\ }\textbf {\bibinfo {volume} {93}},\ \bibinfo
  {pages} {124007} (\bibinfo {year} {2016})}\BibitemShut {NoStop}%
\bibitem [{\citenamefont {Ajith}(2011)}]{Ajith-2011b}%
  \BibitemOpen
  \bibfield  {author} {\bibinfo {author} {\bibfnamefont {P.}~\bibnamefont
  {Ajith}},\ }\href {\doibase 10.1103/PhysRevD.84.084037} {\bibfield  {journal}
  {\bibinfo  {journal} {Phys. Rev. D}\ }\textbf {\bibinfo {volume} {84}},\
  \bibinfo {pages} {084037} (\bibinfo {year} {2011})}\BibitemShut {NoStop}%
\bibitem [{\citenamefont {Shoemaker}(2010)}]{aLIGOZeroDetHighPower}%
  \BibitemOpen
  \bibfield  {author} {\bibinfo {author} {\bibfnamefont {D.}~\bibnamefont
  {Shoemaker}} (\bibinfo {collaboration} {LIGO Scientific Collaboration}),\
  }\href {https://dcc.ligo.org/LIGO-T0900288/public} {\bibfield  {journal}
  {\bibinfo  {journal} {LIGO Document T0900288-v3}\ } (\bibinfo {year}
  {2010})}\BibitemShut {NoStop}%
\bibitem [{\citenamefont {{LIGO Data Analysis Software Working
  Group}}()}]{lalsuite}%
  \BibitemOpen
  \bibfield  {author} {\bibinfo {author} {\bibnamefont {{LIGO Data Analysis
  Software Working Group}}},\ }\href {https://wiki.ligo.org/DASWG/LALSuite}
  {\enquote {\bibinfo {title} {{LALSuite: LSC Algorithm Library Suite}},}\
  }\BibitemShut {NoStop}%
\bibitem [{\citenamefont {Finn}(1992)}]{Finn-1992}%
  \BibitemOpen
  \bibfield  {author} {\bibinfo {author} {\bibfnamefont {L.~S.}\ \bibnamefont
  {Finn}},\ }\href {\doibase 10.1103/PhysRevD.46.5236} {\bibfield  {journal}
  {\bibinfo  {journal} {Phys. Rev. D}\ }\textbf {\bibinfo {volume} {46}},\
  \bibinfo {pages} {5236} (\bibinfo {year} {1992})}\BibitemShut {NoStop}%
\bibitem [{\citenamefont {Sathyaprakash}\ and\ \citenamefont
  {Dhurandhar}(1991)}]{Dhurandhar-1991}%
  \BibitemOpen
  \bibfield  {author} {\bibinfo {author} {\bibfnamefont {B.~S.}\ \bibnamefont
  {Sathyaprakash}}\ and\ \bibinfo {author} {\bibfnamefont {S.~V.}\ \bibnamefont
  {Dhurandhar}},\ }\href {\doibase 10.1103/PhysRevD.44.3819} {\bibfield
  {journal} {\bibinfo  {journal} {Phys. Rev. D}\ }\textbf {\bibinfo {volume}
  {44}},\ \bibinfo {pages} {3819} (\bibinfo {year} {1991})}\BibitemShut
  {NoStop}%
\bibitem [{\citenamefont {Babak}\ \emph {et~al.}(2013)\citenamefont {Babak}
  \emph {et~al.}}]{Babak-2013}%
  \BibitemOpen
  \bibfield  {author} {\bibinfo {author} {\bibfnamefont {S.}~\bibnamefont
  {Babak}} \emph {et~al.},\ }\href {\doibase 10.1103/PhysRevD.87.024033}
  {\bibfield  {journal} {\bibinfo  {journal} {Phys. Rev. D}\ }\textbf {\bibinfo
  {volume} {87}},\ \bibinfo {pages} {024033} (\bibinfo {year}
  {2013})}\BibitemShut {NoStop}%
\bibitem [{\citenamefont {Conway}\ and\ \citenamefont
  {Sloane}(1999)}]{Conway-1999}%
  \BibitemOpen
  \bibfield  {author} {\bibinfo {author} {\bibfnamefont {J.~H.}\ \bibnamefont
  {Conway}}\ and\ \bibinfo {author} {\bibfnamefont {N.~J.~A.}\ \bibnamefont
  {Sloane}},\ }\href {\doibase 10.1007/978-1-4757-6568-7} {\emph {\bibinfo
  {title} {Sphere packings, lattices and groups}}},\ Vol.\ \bibinfo {volume}
  {290}\ (\bibinfo  {publisher} {Springer Science \& Business Media},\ \bibinfo
  {year} {1999})\BibitemShut {NoStop}%
\bibitem [{\citenamefont {Prix}(2007)}]{Prix-2007}%
  \BibitemOpen
  \bibfield  {author} {\bibinfo {author} {\bibfnamefont {R.}~\bibnamefont
  {Prix}},\ }\href {http://stacks.iop.org/0264-9381/24/i=19/a=S11} {\bibfield
  {journal} {\bibinfo  {journal} {Classical and Quantum Gravity}\ }\textbf
  {\bibinfo {volume} {24}},\ \bibinfo {pages} {S481} (\bibinfo {year}
  {2007})}\BibitemShut {NoStop}%
\bibitem [{\citenamefont {Owen}(1996)}]{Owen-1996}%
  \BibitemOpen
  \bibfield  {author} {\bibinfo {author} {\bibfnamefont {B.~J.}\ \bibnamefont
  {Owen}},\ }\href {\doibase 10.1103/PhysRevD.53.6749} {\bibfield  {journal}
  {\bibinfo  {journal} {Phys. Rev. D}\ }\textbf {\bibinfo {volume} {53}},\
  \bibinfo {pages} {6749} (\bibinfo {year} {1996})}\BibitemShut {NoStop}%
\bibitem [{\citenamefont {Owen}\ and\ \citenamefont
  {Sathyaprakash}(1999)}]{Owen-1999}%
  \BibitemOpen
  \bibfield  {author} {\bibinfo {author} {\bibfnamefont {B.~J.}\ \bibnamefont
  {Owen}}\ and\ \bibinfo {author} {\bibfnamefont {B.~S.}\ \bibnamefont
  {Sathyaprakash}},\ }\href {\doibase 10.1103/PhysRevD.60.022002} {\bibfield
  {journal} {\bibinfo  {journal} {Phys. Rev. D}\ }\textbf {\bibinfo {volume}
  {60}},\ \bibinfo {pages} {022002} (\bibinfo {year} {1999})}\BibitemShut
  {NoStop}%
\bibitem [{\citenamefont {Abbott}\ \emph
  {et~al.}(2009{\natexlab{a}})\citenamefont {Abbott} \emph
  {et~al.}}]{psearch1}%
  \BibitemOpen
  \bibfield  {author} {\bibinfo {author} {\bibfnamefont {B.~P.}\ \bibnamefont
  {Abbott}} \emph {et~al.} (\bibinfo {collaboration} {LIGO Scientific
  Collaboration}),\ }\href {\doibase 10.1103/PhysRevD.79.122001} {\bibfield
  {journal} {\bibinfo  {journal} {Phys. Rev. D}\ }\textbf {\bibinfo {volume}
  {79}},\ \bibinfo {pages} {122001} (\bibinfo {year}
  {2009}{\natexlab{a}})}\BibitemShut {NoStop}%
\bibitem [{\citenamefont {Abbott}\ \emph
  {et~al.}(2009{\natexlab{b}})\citenamefont {Abbott} \emph
  {et~al.}}]{psearch2}%
  \BibitemOpen
  \bibfield  {author} {\bibinfo {author} {\bibfnamefont {B.~P.}\ \bibnamefont
  {Abbott}} \emph {et~al.} (\bibinfo {collaboration} {LIGO Scientific
  Collaboration}),\ }\href {\doibase 10.1103/PhysRevD.80.047101} {\bibfield
  {journal} {\bibinfo  {journal} {Phys. Rev. D}\ }\textbf {\bibinfo {volume}
  {80}},\ \bibinfo {pages} {047101} (\bibinfo {year}
  {2009}{\natexlab{b}})}\BibitemShut {NoStop}%
\bibitem [{\citenamefont {Abadie}\ \emph {et~al.}(2010)\citenamefont {Abadie}
  \emph {et~al.}}]{psearch3}%
  \BibitemOpen
  \bibfield  {author} {\bibinfo {author} {\bibfnamefont {J.}~\bibnamefont
  {Abadie}} \emph {et~al.} (\bibinfo {collaboration} {LIGO Scientific
  Collaboration and Virgo Collaboration}),\ }\href {\doibase
  10.1103/PhysRevD.82.102001} {\bibfield  {journal} {\bibinfo  {journal} {Phys.
  Rev. D}\ }\textbf {\bibinfo {volume} {82}},\ \bibinfo {pages} {102001}
  (\bibinfo {year} {2010})}\BibitemShut {NoStop}%
\bibitem [{\citenamefont {Keppel}\ \emph {et~al.}(2013)\citenamefont {Keppel},
  \citenamefont {Lundgren}, \citenamefont {Owen},\ and\ \citenamefont
  {Zhu}}]{Keppel-2013}%
  \BibitemOpen
  \bibfield  {author} {\bibinfo {author} {\bibfnamefont {D.}~\bibnamefont
  {Keppel}}, \bibinfo {author} {\bibfnamefont {A.~P.}\ \bibnamefont
  {Lundgren}}, \bibinfo {author} {\bibfnamefont {B.~J.}\ \bibnamefont {Owen}},
  \ and\ \bibinfo {author} {\bibfnamefont {H.}~\bibnamefont {Zhu}},\ }\href
  {\doibase 10.1103/PhysRevD.88.063002} {\bibfield  {journal} {\bibinfo
  {journal} {Phys. Rev. D}\ }\textbf {\bibinfo {volume} {88}},\ \bibinfo
  {pages} {063002} (\bibinfo {year} {2013})}\BibitemShut {NoStop}%
\bibitem [{\citenamefont {Harry}\ \emph {et~al.}(2014)\citenamefont {Harry},
  \citenamefont {Nitz}, \citenamefont {Brown}, \citenamefont {Lundgren},
  \citenamefont {Ochsner},\ and\ \citenamefont {Keppel}}]{Harry-2014}%
  \BibitemOpen
  \bibfield  {author} {\bibinfo {author} {\bibfnamefont {I.~W.}\ \bibnamefont
  {Harry}}, \bibinfo {author} {\bibfnamefont {A.~H.}\ \bibnamefont {Nitz}},
  \bibinfo {author} {\bibfnamefont {D.~A.}\ \bibnamefont {Brown}}, \bibinfo
  {author} {\bibfnamefont {A.~P.}\ \bibnamefont {Lundgren}}, \bibinfo {author}
  {\bibfnamefont {E.}~\bibnamefont {Ochsner}}, \ and\ \bibinfo {author}
  {\bibfnamefont {D.}~\bibnamefont {Keppel}},\ }\href {\doibase
  10.1103/PhysRevD.89.024010} {\bibfield  {journal} {\bibinfo  {journal} {Phys.
  Rev. D}\ }\textbf {\bibinfo {volume} {89}},\ \bibinfo {pages} {024010}
  (\bibinfo {year} {2014})}\BibitemShut {NoStop}%
\bibitem [{\citenamefont {Pai}\ and\ \citenamefont
  {Arun}(2013)}]{Archana-2013}%
  \BibitemOpen
  \bibfield  {author} {\bibinfo {author} {\bibfnamefont {A.}~\bibnamefont
  {Pai}}\ and\ \bibinfo {author} {\bibfnamefont {K.~G.}\ \bibnamefont {Arun}},\
  }\href {http://stacks.iop.org/0264-9381/30/i=2/a=025011} {\bibfield
  {journal} {\bibinfo  {journal} {Classical and Quantum Gravity}\ }\textbf
  {\bibinfo {volume} {30}},\ \bibinfo {pages} {025011} (\bibinfo {year}
  {2013})}\BibitemShut {NoStop}%
\bibitem [{\citenamefont {Nitz}\ \emph {et~al.}(2016)\citenamefont {Nitz},
  \citenamefont {Harry}, \citenamefont {Biwer}, \citenamefont {Willis},
  \citenamefont {Brown}, \citenamefont {Pekowsky}, \citenamefont {Canton},
  \citenamefont {Dent}, \citenamefont {Williamson}, \citenamefont {Capano},
  \citenamefont {Kumar}, \citenamefont {Lenona}, \citenamefont {De},
  \citenamefont {micamu}, \citenamefont {Fairhurst}, \citenamefont {tjma12},
  \citenamefont {Nielsen}, \citenamefont {Shasvath}, \citenamefont {Babak},
  \citenamefont {Machenschalk}, \citenamefont {Singer}, \citenamefont
  {Macleod}, \citenamefont {Reyes}, \citenamefont {Sugar}, \citenamefont
  {Couvares}, \citenamefont {Bockelman}, \citenamefont {Lundgren},
  \citenamefont {Tewari}, \citenamefont {Ohme},\ and\ \citenamefont
  {Veitch}}]{pycbc}%
  \BibitemOpen
  \bibfield  {author} {\bibinfo {author} {\bibfnamefont {A.}~\bibnamefont
  {Nitz}}, \bibinfo {author} {\bibfnamefont {I.~W.}\ \bibnamefont {Harry}},
  \bibinfo {author} {\bibfnamefont {C.~M.}\ \bibnamefont {Biwer}}, \bibinfo
  {author} {\bibfnamefont {J.}~\bibnamefont {Willis}}, \bibinfo {author}
  {\bibfnamefont {D.}~\bibnamefont {Brown}}, \bibinfo {author} {\bibfnamefont
  {L.}~\bibnamefont {Pekowsky}}, \bibinfo {author} {\bibfnamefont {T.~D.}\
  \bibnamefont {Canton}}, \bibinfo {author} {\bibfnamefont {T.}~\bibnamefont
  {Dent}}, \bibinfo {author} {\bibfnamefont {A.~R.}\ \bibnamefont
  {Williamson}}, \bibinfo {author} {\bibfnamefont {C.}~\bibnamefont {Capano}},
  \bibinfo {author} {\bibfnamefont {P.}~\bibnamefont {Kumar}}, \bibinfo
  {author} {\bibnamefont {Lenona}}, \bibinfo {author} {\bibfnamefont
  {S.}~\bibnamefont {De}}, \bibinfo {author} {\bibnamefont {micamu}}, \bibinfo
  {author} {\bibfnamefont {S.}~\bibnamefont {Fairhurst}}, \bibinfo {author}
  {\bibnamefont {tjma12}}, \bibinfo {author} {\bibfnamefont {A.}~\bibnamefont
  {Nielsen}}, \bibinfo {author} {\bibnamefont {Shasvath}}, \bibinfo {author}
  {\bibfnamefont {S.}~\bibnamefont {Babak}}, \bibinfo {author} {\bibfnamefont
  {B.}~\bibnamefont {Machenschalk}}, \bibinfo {author} {\bibfnamefont
  {L.}~\bibnamefont {Singer}}, \bibinfo {author} {\bibfnamefont
  {D.}~\bibnamefont {Macleod}}, \bibinfo {author} {\bibfnamefont
  {S.}~\bibnamefont {Reyes}}, \bibinfo {author} {\bibfnamefont
  {C.}~\bibnamefont {Sugar}}, \bibinfo {author} {\bibnamefont {Couvares}},
  \bibinfo {author} {\bibfnamefont {B.}~\bibnamefont {Bockelman}}, \bibinfo
  {author} {\bibfnamefont {A.}~\bibnamefont {Lundgren}}, \bibinfo {author}
  {\bibfnamefont {V.}~\bibnamefont {Tewari}}, \bibinfo {author} {\bibfnamefont
  {F.}~\bibnamefont {Ohme}}, \ and\ \bibinfo {author} {\bibfnamefont
  {J.}~\bibnamefont {Veitch}},\ }\href {\doibase 10.5281/zenodo.164940}
  {\enquote {\bibinfo {title} {ligo-cbc/pycbc: Er10 production release 2},}\ }
  (\bibinfo {year} {2016})\BibitemShut {NoStop}%
\bibitem [{\citenamefont {Usman}\ \emph {et~al.}(2016)\citenamefont {Usman}
  \emph {et~al.}}]{Usman-2015}%
  \BibitemOpen
  \bibfield  {author} {\bibinfo {author} {\bibfnamefont {S.~A.}\ \bibnamefont
  {Usman}} \emph {et~al.},\ }\href {\doibase 10.1088/0264-9381/33/21/215004}
  {\bibfield  {journal} {\bibinfo  {journal} {Class. Quant. Grav.}\ }\textbf
  {\bibinfo {volume} {33}},\ \bibinfo {pages} {215004} (\bibinfo {year}
  {2016})},\ \Eprint {http://arxiv.org/abs/1508.02357} {arXiv:1508.02357
  [gr-qc]} \BibitemShut {NoStop}%
%%CITATION = ARXIV:1508.02357;%%
\bibitem [{\citenamefont {Dal~Canton}\ \emph {et~al.}(2014)\citenamefont
  {Dal~Canton} \emph {et~al.}}]{Canton-2014}%
  \BibitemOpen
  \bibfield  {author} {\bibinfo {author} {\bibfnamefont {T.}~\bibnamefont
  {Dal~Canton}} \emph {et~al.},\ }\href {\doibase 10.1103/PhysRevD.90.082004}
  {\bibfield  {journal} {\bibinfo  {journal} {Phys. Rev.}\ }\textbf {\bibinfo
  {volume} {D90}},\ \bibinfo {pages} {082004} (\bibinfo {year} {2014})},\
  \Eprint {http://arxiv.org/abs/1405.6731} {arXiv:1405.6731 [gr-qc]}
  \BibitemShut {NoStop}%
%%CITATION = ARXIV:1405.6731;%%
\bibitem [{\citenamefont {Allen}\ \emph {et~al.}(2012)\citenamefont {Allen},
  \citenamefont {Anderson}, \citenamefont {Brady}, \citenamefont {Brown},\ and\
  \citenamefont {Creighton}}]{findchirp}%
  \BibitemOpen
  \bibfield  {author} {\bibinfo {author} {\bibfnamefont {B.}~\bibnamefont
  {Allen}}, \bibinfo {author} {\bibfnamefont {W.~G.}\ \bibnamefont {Anderson}},
  \bibinfo {author} {\bibfnamefont {P.~R.}\ \bibnamefont {Brady}}, \bibinfo
  {author} {\bibfnamefont {D.~A.}\ \bibnamefont {Brown}}, \ and\ \bibinfo
  {author} {\bibfnamefont {J.~D.~E.}\ \bibnamefont {Creighton}},\ }\href
  {\doibase 10.1103/PhysRevD.85.122006} {\bibfield  {journal} {\bibinfo
  {journal} {Phys. Rev. D}\ }\textbf {\bibinfo {volume} {85}},\ \bibinfo
  {pages} {122006} (\bibinfo {year} {2012})}\BibitemShut {NoStop}%
\bibitem [{\citenamefont {Hibbard}(1962)}]{bst1}%
  \BibitemOpen
  \bibfield  {author} {\bibinfo {author} {\bibfnamefont {T.~N.}\ \bibnamefont
  {Hibbard}},\ }\href {\doibase 10.1145/321105.321108} {\bibfield  {journal}
  {\bibinfo  {journal} {J. ACM}\ }\textbf {\bibinfo {volume} {9}},\ \bibinfo
  {pages} {13} (\bibinfo {year} {1962})}\BibitemShut {NoStop}%
\bibitem [{\citenamefont {Privitera}\ \emph {et~al.}(2014)\citenamefont
  {Privitera}, \citenamefont {Mohapatra}, \citenamefont {Ajith}, \citenamefont
  {Cannon}, \citenamefont {Fotopoulos}, \citenamefont {Frei}, \citenamefont
  {Hanna}, \citenamefont {Weinstein},\ and\ \citenamefont
  {Whelan}}]{Privertia-2014}%
  \BibitemOpen
  \bibfield  {author} {\bibinfo {author} {\bibfnamefont {S.}~\bibnamefont
  {Privitera}}, \bibinfo {author} {\bibfnamefont {S.~R.~P.}\ \bibnamefont
  {Mohapatra}}, \bibinfo {author} {\bibfnamefont {P.}~\bibnamefont {Ajith}},
  \bibinfo {author} {\bibfnamefont {K.}~\bibnamefont {Cannon}}, \bibinfo
  {author} {\bibfnamefont {N.}~\bibnamefont {Fotopoulos}}, \bibinfo {author}
  {\bibfnamefont {M.~A.}\ \bibnamefont {Frei}}, \bibinfo {author}
  {\bibfnamefont {C.}~\bibnamefont {Hanna}}, \bibinfo {author} {\bibfnamefont
  {A.~J.}\ \bibnamefont {Weinstein}}, \ and\ \bibinfo {author} {\bibfnamefont
  {J.~T.}\ \bibnamefont {Whelan}},\ }\href {\doibase
  10.1103/PhysRevD.89.024003} {\bibfield  {journal} {\bibinfo  {journal} {Phys.
  Rev. D}\ }\textbf {\bibinfo {volume} {89}},\ \bibinfo {pages} {024003}
  (\bibinfo {year} {2014})}\BibitemShut {NoStop}%
\bibitem [{\citenamefont {Abbott}\ \emph
  {et~al.}(2016{\natexlab{c}})\citenamefont {Abbott} \emph
  {et~al.}}]{gw15092014:CBC}%
  \BibitemOpen
  \bibfield  {author} {\bibinfo {author} {\bibfnamefont {B.~P.}\ \bibnamefont
  {Abbott}} \emph {et~al.} (\bibinfo {collaboration} {LIGO Scientific
  Collaboration and Virgo Collaboration}),\ }\href {\doibase
  10.1103/PhysRevD.93.122003} {\bibfield  {journal} {\bibinfo  {journal} {Phys.
  Rev. D}\ }\textbf {\bibinfo {volume} {93}},\ \bibinfo {pages} {122003}
  (\bibinfo {year} {2016}{\natexlab{c}})}\BibitemShut {NoStop}%
\bibitem [{\citenamefont {Sch{\"{u}}rmann}\ and\ \citenamefont
  {Vallentin}(2006)}]{Schurmann-2006}%
  \BibitemOpen
  \bibfield  {author} {\bibinfo {author} {\bibfnamefont {A.}~\bibnamefont
  {Sch{\"{u}}rmann}}\ and\ \bibinfo {author} {\bibfnamefont {F.}~\bibnamefont
  {Vallentin}},\ }\href {\doibase 10.1007/s00454-005-1202-2} {\bibfield
  {journal} {\bibinfo  {journal} {Discrete {\&} Computational Geometry}\
  }\textbf {\bibinfo {volume} {35}},\ \bibinfo {pages} {73} (\bibinfo {year}
  {2006})}\BibitemShut {NoStop}%
\bibitem [{\citenamefont {Thomson}(1887)}]{kelvin}%
  \BibitemOpen
  \bibfield  {author} {\bibinfo {author} {\bibfnamefont {S.~W.}\ \bibnamefont
  {Thomson}},\ }\href {\doibase 10.1080/14786448708628135} {\bibfield
  {journal} {\bibinfo  {journal} {Philosophical Magazine Series 5}\ }\textbf
  {\bibinfo {volume} {24}},\ \bibinfo {pages} {503} (\bibinfo {year}
  {1887})}\BibitemShut {NoStop}%
\bibitem [{\citenamefont {Douglas}(1959)}]{bst2}%
  \BibitemOpen
  \bibfield  {author} {\bibinfo {author} {\bibfnamefont {A.~S.}\ \bibnamefont
  {Douglas}},\ }\href {\doibase 10.1093/comjnl/2.1.1} {\bibfield  {journal}
  {\bibinfo  {journal} {The Computer Journal}\ }\textbf {\bibinfo {volume}
  {2}},\ \bibinfo {pages} {1} (\bibinfo {year} {1959})}\BibitemShut {NoStop}%
\bibitem [{\citenamefont {Windley}(1960)}]{bst3}%
  \BibitemOpen
  \bibfield  {author} {\bibinfo {author} {\bibfnamefont {P.~F.}\ \bibnamefont
  {Windley}},\ }\href {\doibase 10.1093/comjnl/3.2.84} {\bibfield  {journal}
  {\bibinfo  {journal} {The Computer Journal}\ }\textbf {\bibinfo {volume}
  {3}},\ \bibinfo {pages} {84} (\bibinfo {year} {1960})}\BibitemShut {NoStop}%
\bibitem [{\citenamefont {Kalaghatgi}\ \emph {et~al.}(2015)\citenamefont
  {Kalaghatgi}, \citenamefont {Ajith},\ and\ \citenamefont
  {Arun}}]{Ajith-2015}%
  \BibitemOpen
  \bibfield  {author} {\bibinfo {author} {\bibfnamefont {C.}~\bibnamefont
  {Kalaghatgi}}, \bibinfo {author} {\bibfnamefont {P.}~\bibnamefont {Ajith}}, \
  and\ \bibinfo {author} {\bibfnamefont {K.~G.}\ \bibnamefont {Arun}},\ }\href
  {\doibase 10.1103/PhysRevD.91.124042} {\bibfield  {journal} {\bibinfo
  {journal} {Phys. Rev. D}\ }\textbf {\bibinfo {volume} {91}},\ \bibinfo
  {pages} {124042} (\bibinfo {year} {2015})}\BibitemShut {NoStop}%
\bibitem [{\citenamefont {Apostolatos}(1995)}]{Apostolatos-1995}%
  \BibitemOpen
  \bibfield  {author} {\bibinfo {author} {\bibfnamefont {T.~A.}\ \bibnamefont
  {Apostolatos}},\ }\href {\doibase 10.1103/PhysRevD.52.605} {\bibfield
  {journal} {\bibinfo  {journal} {Phys. Rev. D}\ }\textbf {\bibinfo {volume}
  {52}},\ \bibinfo {pages} {605} (\bibinfo {year} {1995})}\BibitemShut
  {NoStop}%
\bibitem [{\citenamefont {Khan}\ \emph {et~al.}(2016)\citenamefont {Khan},
  \citenamefont {Husa}, \citenamefont {Hannam}, \citenamefont {Ohme},
  \citenamefont {P\"urrer}, \citenamefont {Forteza},\ and\ \citenamefont
  {Boh\'e}}]{Khan-2016}%
  \BibitemOpen
  \bibfield  {author} {\bibinfo {author} {\bibfnamefont {S.}~\bibnamefont
  {Khan}}, \bibinfo {author} {\bibfnamefont {S.}~\bibnamefont {Husa}}, \bibinfo
  {author} {\bibfnamefont {M.}~\bibnamefont {Hannam}}, \bibinfo {author}
  {\bibfnamefont {F.}~\bibnamefont {Ohme}}, \bibinfo {author} {\bibfnamefont
  {M.}~\bibnamefont {P\"urrer}}, \bibinfo {author} {\bibfnamefont {X.~J.}\
  \bibnamefont {Forteza}}, \ and\ \bibinfo {author} {\bibfnamefont
  {A.}~\bibnamefont {Boh\'e}},\ }\href {\doibase 10.1103/PhysRevD.93.044007}
  {\bibfield  {journal} {\bibinfo  {journal} {Phys. Rev. D}\ }\textbf {\bibinfo
  {volume} {93}},\ \bibinfo {pages} {044007} (\bibinfo {year}
  {2016})}\BibitemShut {NoStop}%
\bibitem [{\citenamefont {Husa}\ \emph {et~al.}(2016)\citenamefont {Husa},
  \citenamefont {Khan}, \citenamefont {Hannam}, \citenamefont {P\"urrer},
  \citenamefont {Ohme}, \citenamefont {Forteza},\ and\ \citenamefont
  {Boh\'e}}]{Husa-2016}%
  \BibitemOpen
  \bibfield  {author} {\bibinfo {author} {\bibfnamefont {S.}~\bibnamefont
  {Husa}}, \bibinfo {author} {\bibfnamefont {S.}~\bibnamefont {Khan}}, \bibinfo
  {author} {\bibfnamefont {M.}~\bibnamefont {Hannam}}, \bibinfo {author}
  {\bibfnamefont {M.}~\bibnamefont {P\"urrer}}, \bibinfo {author}
  {\bibfnamefont {F.}~\bibnamefont {Ohme}}, \bibinfo {author} {\bibfnamefont
  {X.~J.}\ \bibnamefont {Forteza}}, \ and\ \bibinfo {author} {\bibfnamefont
  {A.}~\bibnamefont {Boh\'e}},\ }\href {\doibase 10.1103/PhysRevD.93.044006}
  {\bibfield  {journal} {\bibinfo  {journal} {Phys. Rev. D}\ }\textbf {\bibinfo
  {volume} {93}},\ \bibinfo {pages} {044006} (\bibinfo {year}
  {2016})}\BibitemShut {NoStop}%
\bibitem [{\citenamefont {Taracchini}\ \emph {et~al.}(2014)\citenamefont
  {Taracchini}, \citenamefont {Buonanno}, \citenamefont {Pan}, \citenamefont
  {Hinderer}, \citenamefont {Boyle} \emph {et~al.}}]{Taracchini-2014}%
  \BibitemOpen
  \bibfield  {author} {\bibinfo {author} {\bibfnamefont {A.}~\bibnamefont
  {Taracchini}}, \bibinfo {author} {\bibfnamefont {A.}~\bibnamefont
  {Buonanno}}, \bibinfo {author} {\bibfnamefont {Y.}~\bibnamefont {Pan}},
  \bibinfo {author} {\bibfnamefont {T.}~\bibnamefont {Hinderer}}, \bibinfo
  {author} {\bibfnamefont {M.}~\bibnamefont {Boyle}},  \emph {et~al.},\ }\href
  {\doibase 10.1103/PhysRevD.89.061502} {\bibfield  {journal} {\bibinfo
  {journal} {Phys. Rev. D}\ }\textbf {\bibinfo {volume} {89}},\ \bibinfo
  {pages} {061502} (\bibinfo {year} {2014})}\BibitemShut {NoStop}%
\bibitem [{\citenamefont {P\"urrer}(2014)}]{Purrer-2014}%
  \BibitemOpen
  \bibfield  {author} {\bibinfo {author} {\bibfnamefont {M.}~\bibnamefont
  {P\"urrer}},\ }\href {http://stacks.iop.org/0264-9381/31/i=19/a=195010}
  {\bibfield  {journal} {\bibinfo  {journal} {Classical and Quantum Gravity}\
  }\textbf {\bibinfo {volume} {31}},\ \bibinfo {pages} {195010} (\bibinfo
  {year} {2014})}\BibitemShut {NoStop}%
\bibitem [{\citenamefont {Alam}\ and\ \citenamefont {Haas}(2006)}]{spf}%
  \BibitemOpen
  \bibfield  {author} {\bibinfo {author} {\bibfnamefont {S.~M.~N.}\
  \bibnamefont {Alam}}\ and\ \bibinfo {author} {\bibfnamefont {Z.~J.}\
  \bibnamefont {Haas}},\ }in\ \href {\doibase 10.1145/1161089.1161128} {\emph
  {\bibinfo {booktitle} {Proceedings of the 12th Annual International
  Conference on Mobile Computing and Networking}}},\ \bibinfo {series and
  number} {MobiCom '06}\ (\bibinfo  {publisher} {ACM},\ \bibinfo {address} {New
  York, NY, USA},\ \bibinfo {year} {2006})\ pp.\ \bibinfo {pages}
  {346--357}\BibitemShut {NoStop}%
\end{thebibliography}%
\bibliographystyle{apsrev4-1}

\end{document}